\documentstyle[11pt,a4]{article}

\def\boxx{\vcenter{\vbox{\hrule height.4pt
          \hbox{\vrule width.4pt height8pt
          \kern8pt\vrule width.4pt}\hrule height.4pt}}}

\def\Tr{{\rm Tr}}

\def\mbh{|{\bf B\cdot H}|}
\def\re{{\rm Re\,}}

\title{Infrared behaviour, sources and the Schwinger action principle}
\date{19 May 1994\\(Revised 7th November 1994)}
\author{Mark Burgess}
\begin{document}
\maketitle
\centerline {Institute of Physics, University of Oslo}
\centerline {P.O. BOX 1048, Blindern, 0316 OSLO 3, NORWAY}

\bigskip
\bigskip
\bigskip

\begin{abstract}
An action principle technique is used to examine
the infra-red problem in the effective action for
gauge field theories. It is shown by means of a dynamical analogy
that
the renormalization group and the ansatz of non-local
sources can be simultaneously understood through generalized variations of
an action
supplemented by sources in the manner of the Schwinger action principle.

It is shown that indescriminate resummations of the effective
potential can lead to erroneous conclusions about
phase transitions in a gauge theory if they
correspond to a partial resummation of matter self-energies at
the expense of the gauge sector. The action principle method
illuminates the reason for this and shows a way of proceeding, without
having to go to non-zero momentum.
Some examples are computed to lowest order to compare to results
previously obtained by renormalization group analysis as well as to
prepare for future work.
The utility of the present formalism is a greater
insight into symmetries of the effective action which can be exploited
to regulate the theory perturbatively in the infrared.
The dynamical structure suggests a connection with Chern-Simons theory in
describing the infrared limit of gauge theories.

Finally the dynamical analogy leads to an obvious comparison with
a class of phenomenlogical non-equilibrium lagrangians in
which time-dependent couplings are used to model the influence of
an external system. These models and some of
their implications are discussed in terms of the action principle.
\end{abstract}

\section{Introduction}
Understanding of the massless limit of field theories is essential for
the description of many physical phenomena. Progress has been made in
recent years through a variety of techniques
for resumming the perturbation series and regulating infrared divergences.
Of the many techniques available for dealing with non-perturbative behaviour
in massless theories, two main schools of thought have emerged:
that of ring or daisy summations and that of the renormalization group.
The former represents a method for systematically including a class
of contributions from all orders of conventional perturbation theory
while the latter concerns the redefinition of variables in the theory
so as to most effectively reorganize the conventional perturbation series.

Attention has focussed particularly on a class of theories which undergo
second order (continuous) phase transitions.
In such systems, information about
phase structure and the order of transitions can be obtained directly
by use of field theoretical renormalization group arguments.
Their phases and phase transitions correspond
to scale-invariant points in the parameter space. The success of this program
is partly due to its preoccupation with the simplest scalar field
theories; in particular, it seems that in
systems exhibiting a fluctuation driven first order phase transitions
there are no such perturbatively accessible fixed points
on which to infer the phase structure. An important reason for this
is the very fact that, while the continuous transition is essentially
classical (weak coupling), the discontinuous (first order) transition is an
essentially
quantum (strong coupling) phenomenon and there
is no particular reason why it should
be amenable to normal perturbative methods.

A few authors have pursued a parallel line of development, generalizing
the notion of the effective action to include certain non-local sources
\cite{dahmen1,cornwall1}.
These sources capture some of the non-linear, non-local structure of
the effective action. Here the preoccupation has been with
the bi-local corrections which sum classes of daisy or superdaisy
diagrams in the perturbation expansion, thus the method falls into
the first of the two classes discussed above. The two philosophies
appear conceptually different but can be related if one understands
the effect of a renormalization on the perturbation series.

In the present paper, it is shown that both approaches can be
considered as arising from an application of the Schwinger action
principle\cite{schwinger0,schwinger1} written in such a way that
renormalization group transformations appear as canonical transformations.
The result is a pseudo-dynamical problem which can be straightforwardly
examined and compared to the above alternatives. The principal advantage
of this viewpoint is that
a generalized variation of the action leads to a combined canonical and
renormalization group transformation, providing a renormalization group
improved effective action more directly. Furthermore,
the action principle is a unifying object which reveals the structure
of the renormalization problem in a way which is easily
adapted to problems in curved spacetime\cite{toms1,hu1}, finite
temperature and non-equilibrium
systems\cite{schwinger2}. It also makes a connection with Pad\'e
approximants which in certain regimes can lead to exact
results\cite{nuttall1,graves1}.

\section{The renormalization group as an infrared regulator}

The issue of `resummation'\footnote{The quotes refer to the fact that
this phrase is only intended heuristically, since no
explicit sum is ever performed.}
and renormalization can be illustrated simply when
vertex conditions are used to implement the renormalization procedure.
The vertex method has two formally independent
motivations. First of all, the phenomenological coupling constants
(including the mass) are generally regarded as being given by the
curvatures of the full effective potential, including all radiative
corrections, rather than by the parameters which appear in the unquantized
Lagrangian;
secondly, it is advantageous in the context of perturbation theory
to organize the expansion parameters in such a way that the convergence
of the perturbation series (be it asymptotic or otherwise) be optimal.
The latter remark can be understood quite easily from the following observation
about
the Taylor expansion. Consider a function $f(S)$. Around a point $s$
one may expand the function
\begin{equation}
f(S)=f(s+\Delta s) = f(s) + \frac{df}{ds}\Bigg|_s\;\Delta s
+\frac{1}{2!}\frac{d^2f}{ds^2}\Bigg|_s\;(\Delta s)^2+\ldots\label{eq:1}
\end{equation}
where $\Delta s$ is small in some appropriate sense.
The expansion is clearly not unique, since adding a small counterterm $\delta
s$
\begin{equation}
S=(s+\delta s)+(\Delta s-\delta s)\label{eq:2}
\end{equation}
one has
\begin{eqnarray}
f(S)=f(s+\delta s) + \frac{df}{ds}\Bigg|_{s+\delta s}(\Delta s-\delta s)
+\frac{1}{2!}\frac{d^2f}{ds^2}\Bigg|_{s+\delta s}(\Delta s-\delta s)^2+\ldots
\label{eq:3}\end{eqnarray}
This series is somewhat different. Its convergence properties might be
better or worse than those of (\ref{eq:1}), depending on the magntitude of
$\Delta s - \delta s$. The relationship between the two series may be
found by expanding the new coefficients and expansion parameters in
powers of $\delta s$; it is then seen that the two are identical provided one
works to a consistent order in $\delta s$ and $\Delta s$.
The function of $\delta s$ is to reorganize the size of terms in the series.
In field theory one uses this property
to advantage by defining the renormalized coupling constants as the
coefficients
in the Taylor expansion of the quantum effective potential in powers
of $\overline \phi$.
\begin{equation}
V_{eff}(\overline\phi) = \Lambda +
\sigma\overline\phi+\frac{1}{2}m^2\overline\phi^ 2
+\frac{1}{3}\eta\overline\phi^3+\frac{1}{4!}\lambda\overline\phi^4\ldots
\label{eq:4}\end{equation}
After a redefinition of the coefficients
it is often said that one has `resummed' the
perturbation series since certain contributions have been formally
absorbed from `unknown' higher orders into `known' parameters, by analogy
with the above Taylor series.
In practice only the leading contributions to higher orders
are accessible in field theory.
Moreover, the situation is complicated by the fact that
the Taylor expansion of the
effective potential may not exist around $\overline\phi=0$.
An example is $\lambda\phi^4$ theory in $3+1$-dimensions, where the
coefficient is proportional to $\ln\overline\phi^2$ which diverges in the
limit. In spite of this,
one can use the same argument as before to regularize the Taylor expansion
by rewriting it about a different point and working to a consistent order
in some appropriate expansion parameter.
The invariance of the final
result under reparameterizations is expressed by a `renormalization group'
equation
\begin{equation}
\frac{d}{d(\delta s)} f(S) = 0,\label{eq:5}
\end{equation}
or the appropriate generalization for $n$ counterterms.
The freedom to define the expansion points of the Taylor coefficients
leads to a method of application for the renormalization
group. To use the property to maximal advantage one notices that every
term in the Taylor series can formally be expanded around a different
point, provided they all lie within a radius of the smallness parameter
of each other.
\begin{equation}
f(S) = \sum_{n}^{m-1}\;f_n(\delta s_n)\Delta s_n(\delta s_n)+ {\cal O}(\delta
s^m)\label{eq:6}
\end{equation}
where $\delta s_n\sim\delta s (\forall n)$.
In that case, the difference between the regularized
expansion and the unregularized expansion is formally of higher order
than the consistent order in $\delta s$ to that which one works.
Although this viewpoint is not obviously conventional,
it will become apparent in forthcoming the sections that
such a regularization scheme corresponds
both to an application of the usual field theoretical renormalization group
and to a variation the $n$-point functions
of a theory independently of one another.
In appendices A and B, it is shown explicitly how the method can be
used to regulate infra-red divergences in $\lambda\phi^4$ theory and in
scalar electrodynamics without the need to calculate beta functions.

One of the motivations for such a procedure is to optimize, in some
sense, the asymptotic
perturbation series sufficiently to
determine the type of phase transition a given theory would undergo
from a perturbative calculation.
For the
two examples, it is known that $\lambda\phi^4$ theory has a
second order (continuous) transition, while scalar electrodynamics has a
first order transition. However, the latter result does not follow simply
from the type of renormalized calculation discussed above. Indeed it can
readily be
used to prove that the transition is of second order. This is
disturbing, particularly since essentially
the same type of resummation is regularly
used in the literature in the guise of a diversity of schemes.
The reason for the false conclusion is that the scalar
mass renormalization leads to an effective
regularization of only the scalar sector; the scalar sector dominates
and the gauge sector is neglected. Regrettably there is no
immediately analogous way of generalizing this method to define finite
regulating counterterms
for the gauge sector, owing both to the gauge invariant
structure of the results and the general absence of a background photon field.
One approach which has been used is to define the theory at non-zero
photon momentum. One then finds a renormalization group flow towards an
apparent fixed point at negative infinity\cite{halperin1,lawrie3}.

In view of the preceding argument, it is natural to ask whether one
might construct a dynamical principle for locating the
optimal perturbation series. Can the
problem be solved by finding the minimum of some effective
energy functional? The purpose of the following section is to show that
such a principle can be obtained in a limited sense
by analogy with the Schwinger action
principle. The extent to which such an approach is useful
is naturally limited by the extent to which it can be calculated
in practice, though in the present paper the method also yields interesting
formal comparisons. In the case of gauge theories it illustrates that
the full BRST symmetry might be utilized to improve perturbative calculations.

In what follows it should be borne in mind that even an optimal improvement
of the conventional perturbation series might
not be enough to obtain the desired information. Some results may be
instrinsically non-perturabtive.
Nevertheless, once an appropriate improvement procedure has been
identified, approximations can be made without prejudicing
a special sector of the field space and the advantage of
obtaining a formal expression is that it should be possible to
see when the approximation is valid.
It proves convenient to proceed from the
Schwinger action principle.

\section{Sources, the action principle and renormalization group flow}

Schwinger's action principle is a differential statement of
changes in the transition amplitudes of a theory under
appropriate variations of the
action. The principle states that, given a complete set of states
characterized by $|q^i\rangle$,
the variation of the amplitude or transformation function is given by
\begin{equation}
\delta\;\langle q^j(t')|q^i(t)\rangle = i\langle q^j(t')|\delta
S|q^i(t)\rangle\label{eq:7}
\end{equation}
where $S$ is the action of the system. If the variations are such that
the end points of the variations are fixed, then one has that
\begin{equation}
\delta\;\langle q^j(t')|q^i(t)\rangle = 0\label{eq:8}
\end{equation}
from which one infers both the operator equations of motion and
the quantum equations of motion of
the physical expectation values $\langle S_{,i}\rangle=0$.
The comma denotes the functional derivative of the action with
respect to the field $q^i$.
If the endpoints are not rigid, then the wisdom of the action principle
lies in the statement that the variation of any dynamical operator $F$
is given by
\begin{equation}
\delta F = -i \alpha [F,G]
\end{equation}
where $G$ is the generator of the transformation.
The form of $G$ follows from the
total derivatives incurred on varying the symmetrized action operator.
$\alpha$ is a constant which must be fixed by demanding that the Hamiltonian
operator be the generator of time translations; it depends on the order
of time derivatives in the theory and their symmetry with respect to the field
variables. This statement leads
directly to the fundamental commutation relations of the system.
The reader is referred to refs. \cite{schwinger0, schwinger00,schwinger1}
for a careful exposition of the action principle.

The utility of the action principle becomes
apparent when one introduces external sources which test
the linear response of the theory to small disturbances $J$. In the simplest
instance one lets $S_{tot}=S+Jq$.
Repeated functional differentiation with respect to the source generates
the $n$-point Green functions of the theory
\begin{equation}
\frac{\delta^n\langle q_2|q_1\rangle}{\delta J_{i_1}\cdots\delta J_{i_n}}
== i^n \langle q_2 | T(q^{i_1}\cdots q^{i_n})| q_1\rangle
\label{eq:9}\end{equation}
which implies that
\begin{equation}
\langle q_2(t')|q_1(t)\rangle_J = \langle q_2|T\exp(iJ_i
q^i)|q_1\rangle,\label{eq:10}
\end{equation}
thus the source serves as both a convenient generator of Green functions
as well as a deformant of the transformation function.
Variation of (\ref{eq:7}), including source term, with respect to $q^i$ now
leads to
\begin{equation}
\Gamma_{,i} = \frac{\langle q^{i}\,'|S_{,i}| q^i \rangle } {\langle
q^{i}\,'|q^i\rangle } = - J_i\label{eq:11}
\end{equation}
where $\Gamma$ is the quantum action functional or effective
action\cite{dewitt1}.
A path integral form is readily obtained from (\ref{eq:10}) by writing
\begin{equation}
\langle q^i(t')|q^i(t)\rangle = \exp{iW[J]} =\int d\mu[\Phi]\exp{i \lbrace
S[\Phi] +J\Phi\rbrace}\label{eq:12}
\end{equation}
where $d\mu[\Phi]$ is an appropriate measure.
The effective action is then obtained by writing the field
$\Phi=\overline\phi+\varphi$ and integrating over $\varphi$. The Legendre
transform $\Gamma[\overline\phi] = W[J] -J\overline\phi$ now yields the
effective action satisfying (\ref{eq:11}).

If the variables in the theory act in the presence of an invariance group,
then a change which corresponds to the action of the group leaves the
transformation function unchanged. For instance, using a condensed notation
in which the summation over indices includes an integration over a
corresponding continuous spacetime label,
\begin{eqnarray}
\delta q^i &=& R^i_\alpha[q]\delta\xi^\alpha\label{eq:13}\\
\delta \langle 0|0\rangle &=& \delta \langle 0|\exp(i
J_iR^i_\alpha\delta\xi^\alpha)|0 \rangle = 0.\label{eq:14}
\end{eqnarray}
The parameters $\xi^\alpha$ which generate infinitesimal
transformations represent an
arbitrary symmetry of the system and the combination
$R^i_\alpha[q]\delta\xi^\alpha$ is formally a `counterterm' in the
sense of an infinitesimal change in renormalization.
Defining $R^i_\alpha=\overline R^i_\alpha+R^i_{\alpha,j}q^j$, one has
infinitesimally\cite{dewitt1}
\begin{equation}
\delta \langle 0|0\rangle = iJ_i\delta\xi^\alpha\left[ \overline R^i_\alpha
-iR^i_{\alpha,j}\frac{\delta}{\delta J_i}\right]\langle 0|T\exp(iJ_j q^j)|0
\rangle=0.\label{eq:15}
\end{equation}
The vanishing of the coefficient in this equation leads to an analogous
`renormalization group' equation for the system as we shall see below.

The renormalized coupling constants of a quantum field theory are constant with
respect to position and time in any fundamental theory, but they are
not independent of the renormalizaton scale $\mu$ at which they are
defined.
The effect of a change of $\mu$ is to vary the
values of the renormalized couplings and expectation values in such a way that
the amplitude $\langle 0 +\infty| 0-\infty\rangle$ is left unchanged.
In the contemporary literature a change of $\mu$ is often referred to as
a `flow'. Clearly this is not a physical flow in the sense of a time variation,
but a flow with respect to another parameter which is non-physical.
In any given equilibrium scenario the value of
$\mu$ is fixed by convenience and any measurable is indepedent of $\mu$
when computed exactly. However the perturbation series depends upon the
renormalized values of the couplings since the effect of a renormalization is
to
absorb the effect of certain composite field operators into new effective
couplings, thereby grouping together like-terms up to a shift in the
vacuum energy.

Based on this straightforward observation it is now asserted that
the action principle can be employed to exploit the formal analogy
between the appearence of expectation values of the quantum
field $\Phi^i\rightarrow\varphi^i+\overline\phi^i$ and the arbitrary
redefinition of the couplings in the action
due to renormalization e.g. $m^2 \rightarrow m_R^2+\delta m^2(\mu^2)$.
Such a
procedure exposes a canonical or symplectic structure in the
symmetry\footnote{After submitting this paper for publication
a related preprint\cite{dolan1} appeared
demonstrating this connection nicely from a geometrical viewpoint.
I was also informed of unpublished
work by C. Stephens and D. O'Conner on which
this reference was based\cite{stephens1} and which preceeds this
paper.}. I am grateful to C. Stephens for his subsequent comments on
this paper.

To make the dynamical analogy most explicit one can make the unusual step
of introducing a fictitious set of states, quite separate from the Fock space,
whose only purpose is to rewrite the renormalization group symmetry in a
new form. One then represents the coupling constants
in the usual Lagrangian as fictitious
operators (matrices) which act only on these states in such a way as to
reproduce the renormalized values of the couplings as their eigenvalues.
This is the essence of the dynamical anology described in the next section.

The couplings can be thought of as varying under a renormalization group
transformation in accord with a scale parameter $\mu$
which takes on the role of a `proper time' variable in the manner
of ref. \cite{schwinger3}.
This step is purely artificial and implies no reinterpretation of the
theory -- nor does the introduction of states imply anything other
than an extra formal step. One should not, for example, impose any
probabilistic interpretation on such states, but rather treat them
as fictitious states whose role is rather trivial.
It does however carry with it the implicit assumption that
the variable $\mu$ generates a true symmetry of the system which need not
be the case in an arbitrary renormalization scheme.
In particular we require the symmetry to be a group since it will be
necessary to assume that the combination of infinitesimal transformations
leads to a finite transformation.

The resulting dynamical analogy, aside from being interesting
from a pedagogical perspective, has a useful role
to play in the investigation of phenomenological non-equilibrium
Lagrangians in which the couplings are regarded as depending on
the real time. We shall return to this issue in section 7.

\section{The dynamical analogy}

To facilitate a better understanding of the above remarks,
consider the addition of a source $J$ to the action, in the form
\begin{equation}
S[\Phi]\rightarrow S[\Phi] +\int dV_x\; J\Phi\label{eq:16}
\end{equation}
The classical equations of motion are clearly given by
\begin{equation}
\frac{\delta S}{\delta \Phi} = -J\label{eq:17}
\end{equation}
If the potential described by $S$ has a global minimum corresponding
to the stationary value of the field $\Phi$, then
the role of the source term is to administer a generalized `force'
which shifts the position of the minimum. The potential must be
at least of quadratic order if the field is to be non-degenerate, otherwise
the source simply generates a symmetry transformation on the variables.

The source works by formally varying the quantity in the action which is
conjugate to the field, i.e. the first derivative of the action with
respect the the field variable. If the potential is not infinitely
degenerate in $\Phi$, this will yield an effective change in the
stationary value of $\Phi$ when the equations of motion are solved,
which is equivalent to the claim that $J$ varies the stationary
solutions of $\Phi$ up to a rescaling of the action. It is
incidentally noteworthy
that, in the context of perturbation theory, the introduction of
a source may affect the order of perturbation theory to which one
must work to achieve a certain accuracy.

In passing to the quantum theory the latter remark transfers
to the expectation values of operator `observables'.
It is now advantageous to make the parametric dependence of the couplings
appear in a manner which is analogous to the dynamics of the system.
Let the couplings be represented by operators acting on quasi-states,
as discussed in the previous section.
\begin{equation}
m^2_R \rightarrow m^2(\mu)|\mu\rangle_Q
\end{equation}
This is a purely formal substitution and implies no
physical interpretation. Clearly the fictitious operators only have an
appropriate meaning inside a scalar of the form $\langle
\mu'|m^2|\mu\rangle_Q$.
The field operator plays an important role in mediating between the physical
quantum states of the field and the quasi-states. This is expressed in
the present formalism by defining
the field operator to operate only on the physical states
while its expectation value acts only on the quasi-states. In this way the
two sets decouple at the formal level.

Let the set of all couplings and non-zero expectation values be
denoted by $\lbrace q^i\rbrace$. We shall see below that an
infinitesimal change in the renormalization
point for all $q^i(\mu)$ can be written in the form of a Hamiltonian
flow
\begin{equation}
q^i(\mu) = q^i(0)e^{-i\int_0^\mu d\mu' H_i}\label{h1}.
\end{equation}
We begin by introducing
sources for each of the $q^i$. Define
\begin{equation}
\Delta S = \int_{\mu_1}^{\mu_2} d\mu \;J_i(\mu) q^i(\mu),\label{h2}
\end{equation}
then by analogy with the Schwinger action principle, we may consider a
change in the quasi-amplitude as arising from the action
of a generator $G$ on the end states in the following manner.
\begin{eqnarray}
\delta \langle\mu'|\mu\rangle_Q &=& i \langle\mu'|G'-G|\mu\rangle_Q\label{h3}\\
                    &=&  i \langle\mu'|\int_\mu^{\mu'}
\frac{d}{d\mu}Gd\mu|\mu\rangle_Q\label{h4}
\end{eqnarray}
(\ref{h4}) rests on the tacit assumption that a given $\mu$ generates a
unique set of couplings given appropriate initial conditions -- i.e. that
the renormalization group flows do not cross. There is no general, rigorous
justification for this assumption, except that this is consistent with the
other essential assumption, namely that the symmetry forms a group.
For an infinitesimal transformation of the form (\ref{h1}), we have
\begin{equation}
|q\rangle _Q \rightarrow (1-iH_Q d\mu)|q\rangle _Q
\end{equation}
which, on comparison with (\ref{h3}) implies that
\begin{equation}
H_Q\delta\mu = G = J\delta q.
\end{equation}
Thus infinitesimally we have
\begin{equation}
\frac{\delta}{\delta\mu} \langle\mu'|\mu\rangle _Q =
i\langle\mu'|H_Q|\mu\rangle _Q\label{eq:33}
\end{equation}
where
\begin{equation}
H_Q = \beta^i J_i
\end{equation}
where
\begin{eqnarray}
\beta_i &=& \frac{d q_i(\mu)}{d\mu}\label{eq:31}\\
\gamma_i &=& \beta_i q_i^{-1}\label{eq:30}
\end{eqnarray}
and $H_i = i\gamma_i$ in (\ref{h1}).
The analogy with the proper time method in ref. \cite{schwinger3} is apparent.
The corresponding
`proper time' equations of motion are formally (\ref{eq:33})
and the quasi-amplitude satifies the boundary condition that
\begin{equation}
\lim_{\mu\rightarrow 0}\langle q_i(\mu) | q_j(0)\rangle
= \delta(q_i,q_j)\label{eq:34}.
\end{equation}

We can now proceed to define the generalized effective action by
supplementing the physical states in (\ref{eq:7}) by the quasi-states.
Rather than introducing further new notation, we shall simply refer
to these states by $|q^A\rangle $,
where uppercase roman characters to represent the sum of all
the indices $q^A = \lbrace \Phi^a, q^i\rbrace$, and it will be understood that
this means the
combination of the two sets. Since there is no overlap between the
operators which act on the two types of state, there is no ambiguity in
this procedure.

The generalized action operator including sources is now
\begin{equation}
S[\Phi] \rightarrow S[\Phi] + J_{a}\Phi^a + J_i q^i
\label{eq:18}
\end{equation}
where the repeated indices are assumed to include an integration over
the continuous label: $x$ for indices $a$ and $\mu$ for indices $i$.
The expectation value of the field variable $\overline \phi$ has two
sources as written, but these may be combined to form a single source.

The effective action is now given by
\begin{eqnarray}
\Gamma &=& W[J^i] -  \;\overline q^i J_i -\overline \phi^a J_a\label{eq:19}\\
\overline q^i &=& \langle q^i \rangle = \frac{\delta W}{\delta J_i}
\label{eq:20}\\
\overline \phi^a &=& \langle \phi^a \rangle = \frac{\delta W}{\delta J_a}
\label{eq:20a}
\end{eqnarray}
where the barred quantities represent renormalized values and the part
analogous to the quantum field in the split $q\rightarrow \overline q+\delta q$
is to be eliminated. One therefore expects the present formalism to yield
a set of relations expressing the counterterms in terms of only
the barred variables. Since the additional sources only generate a
symmetry transformation, this apparently modified effective action must
correspond simply to a reparameterization of the usual renormalized
effective action.
Hence, in spite of appearences, $\Gamma$ is not a new object.

Varying (\ref{eq:19}) with respect to $\overline q^i$ involves all the
related generators $R^i_\alpha$. For example, for $\lambda \phi^4$ scalar
field theory, writing explicitly, one has
\begin{eqnarray}
\frac{\delta \Gamma}{\delta \overline \phi} + \frac{\delta \Gamma}{\delta
\overline m^2}
\frac{\delta  \overline m^2}{\delta  \overline \phi} + \frac{\delta
\Gamma}{\delta  \overline \lambda}\frac{\delta \overline \lambda}{\delta
\overline \phi} &=&
\frac{\delta W}{\delta J_\phi} \frac{\delta J_\phi}{\delta \phi}
+ \frac{\delta W}{\delta \overline J_m}\frac{\delta \overline J_m}{\delta
\overline \phi}
+ \frac{\delta W}{\delta J_\lambda}\frac{\delta J_\lambda}{\delta  \overline
\phi}\nonumber\\
&-& \frac{\delta J_\phi}{\delta  \overline \phi} \overline\phi - J_\phi
-\frac{\delta J_m}{\delta  \overline \phi} \overline m^2
- J_m \frac{\delta  \overline m^2}{\delta  \overline \phi}
-\frac{\delta J_\lambda}{\delta  \overline \phi} \overline \lambda
- J_\lambda \frac{\delta \lambda}{\delta  \overline \phi}
\label{eq:21}\end{eqnarray}

\begin{eqnarray}
\frac{\delta \Gamma}{\delta \overline m^2} + \frac{\delta \Gamma}{\delta
\overline \phi}
\frac{\delta  \overline \phi}{\delta  \overline m^2} + \frac{\delta
\Gamma}{\delta  \overline \lambda}\frac{\delta \overline \lambda}{\delta
\overline m^2} &=&
\frac{\delta W}{\delta J_\phi} \frac{\delta J_\phi}{\delta m^2}
+ \frac{\delta W}{\delta J_m}\frac{\delta J_m}{\delta \overline m^2}
+ \frac{\delta W}{\delta J_\lambda}\frac{\delta J_\lambda}{\delta  \overline
m^2}\nonumber\\
&-& \frac{\delta J_\phi}{\delta  \overline m^2} \overline m^2 - J_{m}
-\frac{\delta J_\phi}{\delta  \overline m^2} \overline \phi
- J_\phi \frac{\delta  \overline \phi}{\delta  \overline m^2}
-\frac{\delta J_\lambda}{\delta  \overline m^2} \overline \lambda
- J_\lambda \frac{\delta \lambda}{\delta  \overline m^2}
\label{eq:22}\end{eqnarray}

\begin{eqnarray}
\frac{\delta \Gamma}{\delta \overline \lambda} +
\frac{\delta \Gamma}{\delta \overline m^2}
\frac{\delta  \overline m^2}{\delta  \overline \lambda} +
\frac{\delta \Gamma}{\delta  \overline \phi}
\frac{\delta \overline \phi}{\delta \overline \lambda}
&=&
\frac{\delta W}{\delta J_\phi} \frac{\delta  J_\phi}{\delta \lambda}
+ \frac{\delta W}{\delta J_m}\frac{\delta J_m}{\delta \overline \lambda}
+ \frac{\delta W}{\delta J_\lambda}\frac{\delta J_\lambda}{\delta  \overline
\lambda}\nonumber\\
&-& \frac{\delta J_\lambda}{\delta  \overline \lambda} \overline\lambda
- J_\lambda
-\frac{\delta J_m}{\delta  \overline \lambda} \overline m^2
-  J_m \frac{\delta  \overline m^2}{\delta  \overline \lambda}
-\frac{\delta J_\phi}{\delta  \overline \lambda} \overline \phi
-  J_\phi \frac{\delta \phi}{\delta  \overline \lambda}\label{eq:23}
\end{eqnarray}
leading to the simple equation set
\begin{eqnarray}
\frac{\delta\Gamma}{\delta \overline\phi} &=&  -  J_\phi\label{eq:24}\\
\frac{\delta\Gamma}{\delta \overline m^2} &=& - J_m\label{eq:25}\\
\frac{\delta\Gamma}{\delta \overline \lambda} &=& -J_\lambda
\label{eq:26}\end{eqnarray}
{}From these variational equations it is now straightforward to show that
the action principle is consistent with a `renormalization group' equation
for the system. In particular one sees that the effect of the sources
is to induce a renormalization group transformation on the effective
action.
Consider the case in which the variables $q^i$ are
parametrically dependent on the variable $\mu$.
Then
\begin{equation}
\frac{\delta}{\delta\mu} \langle q^A\,'|q^B\rangle = i\frac{\delta}{\delta\mu}
\langle q^A\,'|S + J_\phi \Phi +J_m m^2 + J_\lambda
\lambda|q^B\rangle\label{eq:27}
\end{equation}
where the sources are not objects of variation.
Integrating the equation
\begin{equation}
\Gamma_{,a} = \langle S_{,a}\rangle\label{eq:28}
\end{equation}
and using the equations of motion for the mean quantities
(\ref{eq:24})-(\ref{eq:26}), one obtains
\begin{equation}
\Bigg\lbrace \frac{\partial}{\partial\mu}-\int dV_x \overline\beta_\phi
\frac{\partial}{\partial\overline\phi}-\overline\beta_m
\frac{\partial}{\partial \overline m^2} - \overline\beta_\lambda
\frac{\partial}{\partial\overline \lambda}  \Bigg\rbrace \Gamma =
\frac{\partial \Lambda(\overline m^2,\mu)}{\partial\mu}\label{eq:29}
\end{equation}
where $\Lambda(\overline m^2,\mu)$ is formally the constant of integration in
(\ref{eq:28}) and $\overline\beta_q = \langle q| \frac{\partial
q}{\partial\mu}|q\rangle_Q$ etc. This
is normally chosen so that $\Gamma[\overline\phi=0]=0$. In
curved spacetime $\Lambda$ acquires the additional significance of being the
cosmological constant. Its presence has an important formal role, though
in practice its only function in flat space is to shift the zero point
energy.

The completeness of the operator picture can be seen by noting
as in \cite{dolan1} that
eqn (\ref{eq:29}) has the form of a Hamilton-Jacobi relation. In the following
relations (\ref{x1}-\ref{x10}) the variable $i$ is not summed over.
\begin{equation}
\frac{\partial \Gamma}{\partial\mu} + H = 0
\end{equation}
where
\begin{equation}
H = -\beta^i\frac{\partial\Gamma}{\partial\overline q^i} = \beta^i J_i = \sum_i
H_i\label{x1}.
\end{equation}
Consistency with the Hamiltonian flow of the couplings is obtained on
noting that
\begin{equation}
H_i = i\beta^i q_i^{-1} = i\beta^i \frac{\partial}{\partial q^i}
\end{equation}
where $q^{-1}$ is the left-inverse of $q^i$.
In particular this gives the quasi-canonical commutation relations
at equal $\mu$. From
\begin{equation}
\frac{\partial}{\partial\mu} q^i = -i [H_i,q^i] = \beta^i
\end{equation}
and, on subsitituting for $H_i$ one obtains
\begin{equation}
[q_i^{-1},q_i] = 1\label{x10}
\end{equation}
or
\begin{equation}
[p_i,q_i] = -i
\end{equation}
where $p_i = i\frac{\partial}{\partial q_i}$, so that the Hamiltonian
projection $H_i = p_i \dot{q^i}$ from which is should be apparent that
$H_i$ is the generator of a quasi-canonical transformation. Finally,
\begin{equation}
\langle p|q\rangle_Q = e^{i\frac{\delta\Gamma}{\delta q}q}.
\end{equation}

An interpretation of this dynamical analogy may be seen in a simple example.
Consider the effective variation of the coupling constant due to
a change in the source. If the parameter space of the
effective action has local maxima
and minima on varying the couplings then they may be located
by considering
\begin{equation}
\frac{\delta\lambda(J_\lambda)}{\delta J_\lambda} = 0\label{eq:35}
\end{equation}
{}From (\ref{eq:20}) this may be written
\begin{equation}
\frac{\delta^2}{\delta J_\lambda} W[J_i] \Bigg|_{J_\lambda} = \langle
\lambda(\mu)\lambda(\mu') \rangle - \langle\lambda(\mu)\rangle
\langle\lambda(\mu')\rangle = 0\label{eq:36}
\end{equation}
which from the equation of motion
\begin{equation}
i\frac{d}{d\mu^2} \langle \lambda(\mu)|\lambda(\mu') \rangle
=\langle \lambda(\mu)|H_\lambda|\lambda(\mu') \rangle\label{eq:37}
\end{equation}
implies that the beta-function $\beta_\lambda$ is vanishing. Thus
the stationary points refer to renormalization group fixed points.
The two pictures are related through a reparameterization, up to
a shift in $\Lambda$.
The dynamical appearence of these equations should not lead to confusion,
$\mu$ is not a true dynamical parameter: it
simply characterizes a change in dependent
variables; the symbolism is intended to inspire obvious analogies, which
we shall find especially significant when turning to deal with with
phenomenological non-equilibrium problems with
time-dependent effective couplings. No {\em physical} evolution is implied.

Although the dynamical analogy is only a formality, it provides us with
one useful trick. Normally one considers variations of the dynamical
variables at a fixed spacetime point. Additional variations of the coordinate
system lead to terms proportional to the energy momentum tensor and
angular momentum. Such variations can be generalized still further to
include a class of variations pertaining to renormalization group
transformations. This equips us with the possibility of obtaining
a so-called renormalization group improved effective action directly
from the action principle. This remark will be amplified in the next
section.

The preoccupation with the renormalization parameter $\mu$ is
no particular
restriction, since the foregoing equations can easily be generalized
to include a dependence of the $q^i$
on a variety of parameters. Recent work on non-equilibrium physics
motivates the
choice of time-dependent couplings and it has been suggested
that such a dependence corresponds to a real-time renormalization group
transformation\cite{hu1,eboli1}.
However, for a system which is not at equilibrium,
such a dynamical change in couplings
will not express a true invariance of the system and thus more
thought must be given to the meaning of such a procedure.
Another invariance parameter which is known to distribute
non-perturbatively is the gauge fixing multiplier. Here the action
principle asserts that the expectation value of the gauge fixing
constraint vanishes in an exact calculation. The Ward identity plays
an analogous role to the renormalization group equation.
For now, we note that
the utility of these formal manipulations lies in ability to
unravel symmetries of the effective action in a dynamical form.
In particular, it gives rise to self-consistent relations between
the Green functions and renormalized variables in the manner of the
gap equation and Schwinger-Dyson equations.
The addition source terms makes themselves felt through the Legendre
transformation in (\ref{eq:18}) and it should be sufficient to
consider these in order to determine the formal expressions for the
counterterms.

An alternative possibility is to consider the present problem in the conjugate
representation in which one attempts to extract the implicit operator
valuedness from the couplings by introducing generators (sources) which
directly resum particular classes of Feynman diagrams.
This is the method of multi-local sources and is discussed further in section
6.

\section{Applications and examples}

A useful feature of the dynamical analogy is that it yields an
indication of the gains and limitations of the renormalization group.
It is particularly useful to know whether a renormalization
of the couplings in the theory is enough to resum all of the required diagrams
in perturbation theory for a given purpose. The purpose of this
section is to use the recursive structure implied by the dynamical
analogy in the preceeding section
to analyse the simplest gauge theory (scalar electrodynamics)
in the massless limit. This model has been examined previously using
more conventional methods\cite{coleman1,halperin1,lawrie3}. The present
formalism confirms previous results with an important corollary and
specifically indicates how one might improve on the simplest calculation.

The central object of interest for many applications is the effective
action. Using the functional evaluation
scheme due to Jackiw\cite{jackiw1}, the effective action may be cast,
for $\lambda\overline\phi^4$ theory,
in the form
\begin{equation}
\Gamma[\overline\phi]= S[\overline \phi]-i \int {\cal D}\varphi \exp i
\lbrace\frac{1}{2}\varphi^a\,S_{,ab}\varphi^b + S_{int}
-\varphi^a\Gamma[\overline\phi]_{,a} -\delta m^2 \frac{\delta\Gamma}{\delta
\overline m^2} -\delta\lambda\frac{\delta\Gamma}{\delta\overline \lambda}
\rbrace\label{eq:38}
\end{equation}
where it is assumed that a background scalar field is present.
The term involving $\Lambda$ can be safely ignored for present purposes
since only variations of the action (potential
differences) have a physical significance.
{}From the effective equations of motion one has
\begin{eqnarray}
\frac{\delta\Gamma}{\delta\overline m^2} &=& \int dV_x \;\frac{1}{2}
\lbrace \overline\phi(x)\overline\phi(x)+ G(x,x)\rbrace
-\delta m^2 \frac{\delta^2\Gamma}{(\delta \overline m^2)^2}
-\delta \lambda \frac{\delta^2\Gamma}{\delta \overline m^2 \delta \overline
\lambda} = 0\label{eq:39} \\
\frac{\delta\Gamma}{\delta\overline\lambda} &=& \int dV_x
\lbrace \frac{1}{4!}\overline\phi(x)
\overline\phi(x)\overline\phi(x)\overline\phi(x) +\frac{1}{4} G(x,x)
\overline\phi(x)\overline\phi(x)\nonumber\\ &+& \frac{1}{8}\overline\phi(x)
G(x,x,x) + \frac{1}{4!} G(x,x,x,x)\rbrace
-\delta q^i\Gamma_{,i}= 0\label{eq:40}
\end{eqnarray}
where, for example,
\begin{eqnarray}
G(x,x') = \frac{\int {\cal D\varphi}\;\varphi(x)\varphi(x')
e^{i\lbrace\frac{1}{2}\varphi^a\,S_{,ab}\varphi^b
+S_{int} - q^i\Gamma,i\rbrace}}{\int {\cal D\varphi}
e^{i\lbrace\frac{1}{2}\varphi^a\,S_{,ab}\varphi^b
+S_{int} - q^i\Gamma,i\rbrace}}\label{eq:41}
\end{eqnarray}
and there are a number of relations of the form
\begin{eqnarray}
\frac{\delta^2\Gamma}{(\delta\overline m^2)^2} &=& \int dV_x\; \frac{1}{2}
\frac{\delta G(x,x)}{\delta\overline m^2} +\cdots\label{eq:42}\\
\frac{\delta^2\Gamma}{\delta\overline m^2\delta\overline\lambda}
&=& \int dV_x \lbrace \frac{1}{4}\overline\phi(x)\overline\phi(x)
\frac{\delta G(x,x)}{\delta\overline m^2} + \frac{1}{4}
\frac{\delta G(x,x,x,x)}{\delta \overline m^2}\rbrace\label{eq:43}\\
\frac{\delta G(x,x)}{\delta\overline m^2} &=& \frac{1}{2}
\int dV_{x'} \lbrace G(x,x,x',x') - G(x,x)G(x',x')\rbrace\label{eq:44}
\end{eqnarray}
Using these expressions in the equations of motion for vanishing
quantum source part, one obtains iterative
differential equations which can be solved for
and used to eliminate $\delta q_i$; to leading order in the loop
counting parameter (which is set to $1$ in this paper) this can
be separated
\begin{eqnarray}
\int \frac{\delta G}{G(x,x)} &=& \int \frac{\delta(\delta m^2)}{\delta
m^2}\label{eq:45}\\
\int \frac{\delta G}{G(x,x,x,x)} &=& \int \frac{\delta(\delta \lambda)}{\delta
\lambda}.\label{eq:46}
\end{eqnarray}
To zeroth order one verifies only the `classical equation of motion'
for the couplings. The remaining
conditions relate therefore to the perturbative corrections
to the standard effective action -- in other words
fluctuations in the true dynamical fields (not the quasi-operators).
They therefore correspond, in a limited sense, to a minimization of
perturbative
corrections, or an optimal perturbation series.
With the partial boundary condition that $\delta m^2 = 0$, $\delta\lambda = 0$
when $\lambda=0$, one may write on dimensional grounds
\begin{eqnarray}
\delta m^2 &=& \lambda G(x,x) \xi\label{eq:47}\\
\delta \lambda &=& \lambda^2 \Tr{G(x,x,x,x)}\xi'\label{eq:48}
\end{eqnarray}
where $\xi,\xi'$ are dimensionless constants. These forms make the
source terms in the action into an invariant form.
These arbitary constants reflect the residual non
uniqueness of the coupling-counterterm split and must therefore
still be chosen by some additional condition. This is to be
expected since we have so far imposed no renormalization conditions
on the parameters of the theory.

The coincidence limit of the Green functions
$G(x,x') \rightarrow G(x,x)$ is
to be regarded in a formal sense, since
in reality this may diverge\footnote{In principle one might absorb the
ultraviolet divergences
by redefinition of the source terms. Here I am ignoring the
ultraviolet behaviour which may be dealt with by minimal subtraction,
for instance,
and treating only the infra-red
behaviour in the present scheme.}.
These expressions can be compared with the usual vertex conditions used
in the method of renormalization by `oversubtraction'. Differentiating
(\ref{eq:38}) in the zero momentum limit leads to
\begin{eqnarray}
\frac{1}{\Omega}\frac{\partial^2\Gamma}{\partial\overline \phi^2}\Bigg
|_{\overline\phi=0}
&=& m^2 + \frac{\lambda}{2}\; G(x,x)\label{eq:49}\\
\frac{1}{\Omega}\frac{\partial^4\Gamma}{\partial\overline \phi^4}\Bigg
|_{\overline\phi=0}
&=& \lambda +\frac{3\lambda^2}{2}\;\Tr G(x,x,x,x)\label{eq:50}
\end{eqnarray}
where only the leading order corrections are included.
$\Omega$ is a spacetime volume scale.
It is seen the that the introduction of the sources generates a term
precisely of the form of that due to a more conventional renormalization
method. Moreover, it is now possible
to see why the approximate
regularization prescription in appendix A has the
desired effect. The arbitrary renormalization point $\Phi_m^2$ plays
the role of a crude approximation to $G(x,x)$. In that particular
calculation, no great accuracy was required, only the presence of a
regulating contribution from $G(x,x)$. In this example nothing is gained
from the new approach, except perhaps the satisfaction of knowing that
an old result can be shown in a new way. It is more interesting to
consider a gauge theory.

The inclusion of gauge fields presents new subtleties
for the renormalization approach, the first of which
arises from the non-linear
dependence of the action on the gauge coupling
in a second order derivative theory -- even
at the classical level. Given the gauge-covariant derivative defined by
$D_\mu=\partial_\mu+ieA_\mu$, one considers an action of the form
\begin{equation}
S = \int dV_x \Big\lbrace (D^\mu\Phi)^\dagger(D_\mu\Phi) + V(\Phi)
\Big\rbrace.\label{eq:51}
\end{equation}
In the quantized theory one is forced to break the invariance group
so as to count only physical fields. Also, since the
gauge parameter distributes non-perturbatively in the usual definition
of the effective action, the specific choice of gauge will
effectively lead to different perturbative expansions, some of which
may be better than others.
In a functional integral representation, this is implemented by a gauge fixing
condition and possibly the introduction of ghost terms\footnote{While
the introduction of ghosts is not a necessity at one loop, practical
calculations in most gauges will demand their introduction at
higher loops. Moreover, the widespread belief that the Jacobian contribution is
unity in scalar electrodynamics is mistaken in general\cite{fradkin1}.},
which must be taken into account by the renormalization. A particular
example is adopted for the remainder of this section, namely
scalar electrodynamics in $3+1$-dimensions, as described in appendix B.
The addition of a source term $J_e e$ now varies the conjugate quantity
\begin{eqnarray}
\frac{\delta\Gamma}{\delta\overline e} = - J_e &=& J^\mu \overline
A_\mu
+ 2e\overline\phi^2\langle A^\mu A_\mu\rangle +2e\overline\phi^2
\langle \overline\eta \eta \rangle
+2e\overline\phi_a\langle \varphi_a\overline \eta \eta \rangle
+ 2e\overline A^\mu \langle A_\mu\varphi_a\varphi_a\rangle\nonumber\\
&+&\frac{1}{\alpha}e\overline\phi^2\langle\varphi_a\varphi_a\rangle
+ 2e\overline\phi_a\langle\varphi_a A^\mu A_\mu\rangle
+e\langle A^\mu A_\mu \varphi_a\varphi_a\rangle - \delta e^i
\frac{\delta^2\Gamma}{\delta\overline e^2}\label{eq:52}
\end{eqnarray}
where
\begin{equation}
J^\mu \equiv i[\Phi(D^\mu\Phi)^\dagger - \Phi^\dagger(D^\mu\Phi)].\label{eq:53}
\end{equation}
and the expectation symbols refer to the exact correlators, defined
in terms of the full action. These contributions arise specifically
from the vacuum polarization in the theory, as would be expected.
The renormalization of the gauge theory is seen to be a complicated
non-linear problem and it is apparent that the full BRST symmetry might be
utilized to generate an effective resummation.

The determination of the counterterm proceeds as before.
Further approximation is now required owing
to the complexity of the gauge symmetry.
Earlier work using the renormalization group calculated by traditional
methods\cite{halperin1,lawrie3} shows that a charge renormalization is
necessary in
order to reveal the possibility of a first order phase transition. This
is generated by renormalizing away from zero momentum in order to
compute a beta-function for the electric charge.
Here we shall attempt to investigate this possibility more directly using
the action principle and the structure generated by the sources.

Information about the phase transitions of the theory
in contained in the infrared limit of the effective action. The extraction
of this information presupposes that the infrared limit is
well behaved. This is not the case in the usual perturbation expansion
since one has to be able to determine the form of the effective potential
for the background $\overline\phi$ field in the limit as one approaches
the origin. The perturbative expression for the effective potential does not
exist at the origin and thus it
must be regulated by some appropriate resummation.

The regulated form of the effective potential around the origin determines the
order of the transition in the following way. In a second order phase
transition,
the addition of a small mass to the $\overline m^2=0$ potential makes the
potential curve monotonically upward implying a single minimum at
the origin. In a first
order transition a minimum away from $\overline\phi=0$ at $\overline m^2=0$
survives for some finite mass correction and the potential curves
downwards at some small value of $\overline\phi$ below the minimum.
One is therefore interested in regulating the infrared behaviour of the
effective potential, namely the small $\overline\phi$ regime.

It ought to be remarked at this juncture that the non-convex form of
the effective potential is a perturbative artefact\cite{wiedemann1} which
has led to much confusion in the past. The present arguments relate
primarily to the approach to phase transitions through perturbation
theory on which the majority of the literature is based. It is understood
that one is working at the edge of perturbative credability here.
The distinction between first and second order is in whether the potential
curves monotonically upward for an infinitesimal mass
perturbation\cite{lawrie1}.

To obtain an infrared regularization one would clearly
like a mass term in the various functional
determinants to supplement $e^2\overline\phi^2$
in the limit of vanishing $\overline\phi$.
To determine whether such a mass can be induced by a renormalization group
transformation, one can follow the formalism of the sources and determine
the appropriate behaviour from the definition of the perturbative
propagator -- which must be regarded as a small quantity if the
diagrammatic expansion is to make sense.
For the scalar field this was
straightfowardly obtained by using the trick in appendix A for renormalizing
the mass. However in the case of the gauge field no direct analogue of such
a mass term exists in the classical Lagrangian and thus there is no source for
such a mass. It is interesting however that, in the present case,
a charge renormalization is partially successful and this
explains why the renormalization group\cite{lawrie3} argument enables
the infrared behavour to be regulated in an asymptotic limit.
The regulating mass arises from vacuum polarization.

Since the gauge field mass is given by $e^2\overline\phi^2$,
one is interested in the behaviour
of the theory approaching the symmetric phase ($\overline\phi\rightarrow 0$)
where perturbation theory is weakest.
\begin{equation}
\frac{\delta^2\Gamma}{\delta e^2} \sim \langle A^\mu
A_\mu\varphi_a\varphi_a\rangle + e \frac{\delta \langle A^\mu
A_\mu\varphi_a\varphi_a\rangle}{\delta(\delta e)}.\label{eq:54}
\end{equation}
In the neighbourhood of $\overline\phi=0$ one has
\begin{equation}
\frac{\delta\Gamma}{\delta\overline e}\Bigg|_{\overline\Phi,\overline A_\mu=0}
\sim (\overline e+\delta e)\langle A^\mu A_\mu \varphi_a\varphi_a\rangle
-\delta e\Bigg\lbrace \langle A^\mu A_\mu \varphi_a\varphi_a\rangle
+e\frac{\delta \langle A^\mu A_\mu \varphi_a\varphi_a\rangle}{\delta(\delta
e)}\Bigg\rbrace=0\label{eq:55}
\end{equation}
for the first order corrections to $\Gamma$,
and one is thus led to the following formal
equation for $\delta e$
\begin{equation}
\int \frac{\overline e \delta(\delta e)}{\delta e(\overline e+\delta e)}
=\frac{\delta\langle A^\mu A_\mu \varphi_a\varphi_a\rangle}{\langle  A^\mu
A_\mu \varphi_a\varphi_a\rangle}.\label{eq:56}
\end{equation}
Expressing this in partial fractions and integrating leads to the
approximate result
\begin{equation}
\delta e \sim e^3 \Tr\langle A^\mu A_\mu \varphi_a\varphi_a\rangle
\xi'\label{eq:57}
\end{equation}
for some $\xi'$. The action principle leads therefore to a leading
order gauge mass of the form
\begin{equation}
e^2\overline\phi^2 \sim \overline e^2\overline\phi^2 +2\overline e\delta e
\overline\phi^2 +\delta e^2\overline\phi^2.\label{eq:58}
\end{equation}
The regularization of the effective potential near $\overline\phi\rightarrow 0$
requires a solution of the form $\delta e\sim \overline\phi^{-1}$, securing
finiteness in the limit. To investigate
the possibility for this one may deduce the limiting behaviour
of the perturbative Green functions from their definitions after a
renormalization group transformation of the kind implied by the
sources.

To lowest order in perturbation theory
one may write $\langle A^\mu A_\mu \varphi_a\varphi_a\rangle\sim
\langle A^\mu A_\mu\rangle\langle\varphi_a\varphi_a\rangle$
thus, to a first approximation, we must obtain self-consistent forms
for the scalar propagator $\Delta$ and the gauge propagator $G$
in the coincidence limit,
which incorporate the appropriate counterterms generated by the sources.
The scalar propagator may be written
\begin{equation}
\Delta = \int \frac{d^nk}{(2\pi)^n}(k^2+\frac{\lambda}{2}
\overline\phi^2+\lambda\xi\Delta)^{-1}
\end{equation}
from which one obtains, using dimensional regularization
\begin{equation}
\Delta = \overline\lambda(\frac{1}{2}\overline\phi^2+\xi\Delta)
\ln \left(
\frac{\lambda(\frac{1}{2}\overline\phi^2+\xi\Delta)}{4\pi\mu^2}\right)
-\frac{\lambda(\frac{1}{2}\overline\phi^2+\xi\Delta)}{4\pi}
\end{equation}
To regulate the divergences as $\overline\phi^2\rightarrow 0$ it is
tempting to let $\mu^2 \sim \frac{1}{2}\lambda\overline\phi^2$
reproducing the procedure in appendix A. One then finds that
$\Delta \sim \lambda\overline\phi^2$ as the background field
vanishes, in accordance with the appendix. However in the gauge
theory the same procedure does not regulate all the divergences and
one is forced to adopt a different tactic. If one holds $\mu^2$ fixed
and considers the $\overline\phi\rightarrow 0$ limit of the
propagator, it is found that
\begin{equation}
\Delta \rightarrow \frac{4\pi\mu^2}{\lambda\xi}\label{EQ:1}.
\end{equation}
This result will be used shortly. The gauge propagator in the
Landau-deWitt gauge may be written
\begin{equation}
G=G^\mu_\mu(x,x) = \int \frac{d^nk}{(2\pi)^n}
\frac{(n-1)}{k^2+e^2\overline\phi^2}.
\end{equation}
Using the form in equation (\ref{eq:58}) and employing dimensional
regularization one obtains the following implict equation for $G$
\begin{equation}
4\pi G=(\overline e^2\overline\phi^2+\overline e^4\overline\phi^2
G\Delta\Omega\xi)
\left(1+3\ln\left(\frac{\overline e^2\overline\phi^2+\overline
e^4\overline\phi^2 G\Delta\Omega\xi}{4\pi\mu^2}\right)\right)
\label{EQ:2}
\end{equation}
where $\Omega$ is a spacetime volume scale. If one uses the form for
$\Delta$ from appendix A, i.e. $\Delta \sim \overline\lambda\overline\phi^2$,
and the implied behaviour of $\mu^2$, then it appears that one can
regulate the infrared divergences in precisely the procedure analogous
to appendix B. The argument in appendix B yields a second order
phase transition, thus the present procedure also indicates a second
order transition, since the curvature of the effective potential
appears to curve upward even for $\overline m^2=0$.
However in the steps above one is forced to make the implicit
assumption that $\overline\lambda\sim\overline e^2$ in order to
consistently avoid large logarithmic corrections in $G$ and it is
known, from Coleman and Weinberg's original argument\cite{coleman1}
that the one-loop calculation in appendix B is not consistent if this
is the case. Thus what the action principle reveals here if this
inconsistency in making a thoughtless resummation of the scalar sector.

One may now proceed according to the second possibility (\ref{EQ:1}) above.
Considering the $\overline\phi\rightarrow 0$ limit of (\ref{EQ:2}), one
obtains the approximate form
\begin{equation}
\exp \left\lbrack \frac{4\pi G}{\overline e^4 \overline\phi^2
G\Delta\xi'\Omega} \right\rbrack
=\frac{\overline e^4\overline\phi^2 G\Delta \xi'\Omega}{4\pi\mu^2}
\end{equation}
and on substituting the asymptotic form of the scalar propagator one
obtains
\begin{equation}
\exp\left\lbrack \frac{\lambda\xi}{\overline e^4
\overline\phi^2\mu^2\xi'\Omega}\right\rbrack
=\frac{\overline e^4\overline\phi^2 G\xi' \Omega}{\lambda \xi}\label{EQ:3}.
\end{equation}
One may now proceed in one of two ways. The first method parallels the
conventional renormalization group approach in refs \cite{halperin1,lawrie3}.
Here we note that, as $\overline\phi\rightarrow 0$, $\delta e$ must
become large in order to prevent the vanishing of the gauge propagator
mass. However, since the effective action is invariant under
the placement of such a counterterm
\begin{equation}
\delta \Gamma = -J_m\delta m^2 -J_e\delta e-J_\lambda\delta\lambda \sim 0
\end{equation}
where $\sim$ indicates that the result is true up to a shift in the
zero point energy. Since both $\delta m^2$ and $\delta e$ must be
positive to avoid sicknesses in the theory, one has that
$\delta \lambda < 0$, since $\Gamma$ varies
approximately symmetrically with
respect to these quantities. If the initial value
of $\overline\lambda$ is sufficiently small, it is thus possible that
$\lambda$ itself can become negative. This is indeed what happens by
conventional methods and one sees from (\ref{EQ:3}) that this is
infact necessary if this equation is to be satisfied as $\overline\phi$
vanishes. However, it is noteworthy that the equation is only
satisfied asymptotically as $\overline\phi$ reaches zero owing
to the exponential behaviour. This parallels the asymptotic fixed
point at $\lambda = -\infty$ in refs \cite{halperin1,lawrie3}.
The fact that $\lambda$ turns negative shows that the
effective action has negative curvature at $\overline m^2=0$,
$\overline\phi=0$ and this is strongly suggestive of a first order
phase transition. It is noteworthy, however, that this conclusion
is reached by desperate means -- as an asymptotic regularization
of infrared divergences, thus
the conclusion rests of the very edge of what one might hope to extract
from perturbation theory.

The second method of proceeding from (\ref{EQ:3}) is to attempt to
cure infrared divergences by finding a solution of the form
\begin{equation}
\frac{\Omega\overline\phi^2 G\xi'}{\xi} = c\label{EQ:4}
\end{equation}
for some constant $c$ which is assumed to be of order unity. Then
one obtains the limiting form
\begin{equation}
\exp \left\lbrack \frac{\overline\lambda G}{\overline e^4 c\mu^2}\right\rbrack
=\frac{\overline e^4}{\overline\lambda}c.
\end{equation}
This may be solved by taking $\overline \lambda \sim \overline e^4$, which
provides a motivation for this choice in Coleman and Weinberg's argument.
It is known that this choice leads to a reliable pertubation theory
which indicates a first order transition. One must again be cautious
however. In writing (\ref{EQ:4}) one is stretching the privelige
to redefine the free parameters $\xi$ to limit of credability and thus
this result is no more trustworthy than that obtained
from the first approach.

The procedure adopted here has been to make the minimally acceptable
approximation. It is natural to wonder whether a more accurate
representation of $\delta e$ would lead to a better result.
Would it, for example, lead to the appearence of a fixed point associate
with a first order phase transition? Examining the form of (\ref{eq:52})
it would appear that any improvement would be difficult to obtain unless
$\overline A_\mu\not=0$. One way to understand this could be that the
phase transition is on the very edge between first and second order for
$\overline A_\mu=0$, but that it is first order
for $\overline A_\mu\not=0$, i.e.
with an electromagnetic background field -- since the present conclusion of
fluctuation induced symmetry breaking can be potentially more reliably
inferred in this case. One might attempt to consider the present theory
as the $\overline A_\mu\rightarrow 0$ limit of the same theory in an
external field. This is essentially the procedure which must be adopted
in \cite{halperin1,lawrie3} since there is no charge renormalization
according to usual regularization schemes unless $\overline A_\mu\not=0$.
Either way, it appears that the phase transition is
weakly first order. No more reliable conclusion has thus far been
obtained by other means to the present author's knowledge.

The conclusion of the above pragmatical approach
via the action principle is not only that familiar renormalization group
results can be reproduced by considering the self-consistent
properties of the propagators, but that the reason for these
results can be seen clearly in the way in which the different Green
functions contribute to the calculation in the infrared limit.

As a further illustration, the present method can be used to calculate the
magnetic
vacuum polarization energy in a non-Abelian gauge theory. This
calculation is known to be perturbatively unreliable and leads to the
so-called Nielsen-Oleson `instability'\cite{nielsen1}, which consists of an
imaginary
contribution to the energy and hence vacuum decay in an external
Abelianized magnetic field. In constrast to the
imposition of a constant electric field, one would not expect
magnetically induced vacuum decay since the magnetic field does no
work on particles. The only apparent alternative is that the imaginary
part is an infrared artifact. The previous computations and
conventions
of reference\cite{burgess3} can be used to obtain an improved one-loop
result which includes an estimate of leading order non-perturbative
corrections for an arbitrary semi-simple gauge group.
Consider the action
\begin{equation}
S_{YM}[A]={{1}\over{4I_2(G_{adj})}}
\int dv_x{\rm Tr}\Big(F^{\mu\nu}F_{\mu\nu}\Big)\;,\label{eq:64}
\end{equation}
for Dynkin index $I_2$ in the Hermitian adjoint representation.
\begin{equation}
F_{\mu\nu}=\partial_\mu A_\nu-\partial_\nu A_\mu+ig[A_\mu,A_\nu]\label{eq:65}
\end{equation}
The gauge-fixing term is
\begin{equation}
          S_{GF}={{1}\over{2\alpha I_2(G_{adj})}}\int dv_x{\rm Tr}\Big({
D}_\mu A^\mu\Big)^2\label{eq:66},
\end{equation}
where
\begin{equation}
{D}_\mu A^\mu=\partial_\mu A^\mu+ig[{\overline A}_\mu,A^\mu]\label{eq:67}
\end{equation}
and the ghost action is
\begin{equation}
          S_{GH}=\int dv_x{\overline \eta}\Big[-{D}^2-gA^\mu
          {D}_\mu-g({D}_\mu A^\mu)\Big]\eta.\label{eq:68}
\end{equation}
A $d+2$ dimensional
Euclidean metric is used. The one-loop operator in the Feynman gauge
is given by
\begin{equation}
\Gamma^{(1)}= -\frac{1}{2}{\rm Tr}\ln(-D^2\delta^{^\mu}_{~\nu}+2igF^\mu_{~\nu})
\label{eq:69}\end{equation}
and the magnetic background field $\overline A_\mu$ is a linear
combination of the generators of the Cartan subalgebra ${\bf B\cdot H}$.
The source of the imaginary part can be seen by replacing the quadratic
derivative operator by its eigenvalues
\begin{equation}
-D^2 \rightarrow p^2 + 2g\mbh(n + 1/2)~~,~~{\rm degeneracy \;per\; unit\;
volume \;\;}\label{eq:70}
{g\mbh\over 2 \pi}
\end{equation}
Note that $D^2$ is a diagonal matrix in the Lie
algebra and the eigenvalues of $F^\mu_{~\nu}$ are straightforwardly
\begin{equation}
2i F^\mu_{~\nu} \rightarrow (+2g\mbh,~~-2g\mbh,~~0\ldots 0)\label{eq:71}
\end{equation}
where $0$ occurs with degeneracy $d$, the one loop contribution is
\begin{eqnarray}
\Gamma^{(1)}_{YM} = - {1\over 4\pi} \sum_{n=0}^\infty \int {d^dp\over (2\pi)^d}
[ d\,
\Tr g\mbh\,
\ln (p^2 + 2 g\mbh (n + 1/2))\nonumber\\
+ \Tr g\mbh\, \ln (p^2 + 2 g\mbh (n + 3/2))\nonumber\\
+ \Tr g\mbh\, \ln (p^2 + 2 g\mbh (n - 1/2))].\label{eq:72}
\end{eqnarray}
When $p=0$, $n=0$ the last logarithm suffers a negative argument.
A quantum mass counterterm of the order $g\mbh$ would eliminate this problem.
Since the non-Abelian theory in non-linear in $A_\mu$, it is possible to
determine that such a mass term exists from the
self-interactions and does not contradict the
structure of the theory. Whilst the classical gauge mass necessarily vanishes
for gauge invariance, a finite mass counterterm is allowed provided it only
resums existing parts of the theory. Note that a charge $g$ renormalization
can never change the minus sign in (\ref{eq:71}) into a plus sign, so a
straightforward renormalization group solution for this does not exist.
The variation
of the effective action with respect to the postulated gauge mass leads to
\begin{equation}
\delta m^2_{YM} = \frac{1}{2}\Tr\langle A^{\mu\,a} A_\mu^a\rangle
\xi\label{eq:73}
\end{equation}
analogously to (\ref{eq:47}). Factors of the Dynkin invariants cancel owing to
the normalization in (\ref{eq:66}). The calculation of the value of this
counterterm
is extremely difficult. An estimate only can be obtained from the results
in ref.\cite{burgess3}. Considering only orders of magnitude, one has
after some calculation
\begin{eqnarray}
\Gamma_{YM}^{(1)} = (4\pi)^2\Tr (2g\mbh)^2\Big\lbrace
-\ln(2g\mbh)\zeta_H(-1,\rho) +\zeta_H(-1,\rho)+\zeta_H(-2,\rho)
\Big\rbrace,\label{eq:74}
\end{eqnarray}
where $\zeta_H(a,b)$ is the Hurwitz zeta function and $\rho = \Tr \delta m^2
/(2g\mbh)$. Assuming $\rho \le 1$ and using the formula
\begin{equation}
\zeta_H (-s, \nu) = {2 \Gamma(s + 1)\over (2\pi)^{s + 1}} \sum_{n=1}^\infty
{\sin (2\pi n \nu - s \pi/2)\over n^{s+1}}~~,~~\re(s)> 0~~,~~0 < \nu \leq 1
\label{eq:75}\end{equation}
one may write $\delta m^2_{YM}\sim \frac{\partial \Gamma}{\partial(\delta
m^2_{YM})}$, whose leading order approximation leads to
\begin{equation}
\delta m^2_{YM} \sim \frac{-4g\mbh(4\pi)^2
+
\sqrt{8(4\pi)^4(2g\mbh)^2+8(2g\mbh)^2(\ln(\frac{2g\mbh}{\mu^2})-\frac{1}{2})^2}}{2(\ln(\frac{2g\mbh}{\mu^2})-\frac{1}{2})}.\label{eq:76}
\end{equation}
In the region of validity of this crude approximation one has
\begin{equation}
\delta m^2 = g\mbh\, \xi.\label{eq:77}
\end{equation}
Since this estimate arises from a perturbative calculation there is naturally
a dependence of the arbitrary scale $\mu^2$, which is transferred above
to the parameter $\xi$. There is freedom within the bounds
of the approximations to choose the value of this parameter. In the
present instance a value of $\xi=1$ is desirable. This corresponds
roughly to logarithms $\ln(\frac{2g\mbh}{\mu^2})$ of the order $(4\pi)^2$
which lies very far outside of the range of one-loop perturbation theory.
The best estimate in a good perturbative regime is $\xi\sim 0.1$ which
is hardly adequate.
However, a larger value does not {\em necessarily} cause a problem for
(\ref{eq:77}) which is
an order of magnitude estimate,
obtained by self-consistently including a subset of higher order contributions.
One can however only
conclude in this instance that it is plausible that the `unstable mode'
is an infrared artifact. Its eventual cancellation
in mean-field perturbation theory requires
presently impossible feats of calculation.
The inclusion of correlations in the self-interacting gauge field is
nonetheless essential, as the action principle indicates.

As a final corollary to the dynamical analogy, it is interesting to
speculate on the role of the Chern-Simons action in the infrared
limit of field theory. Recent work shows that a Chern-Simons term
involving a fictitious gauge field can be used to effect so-called
hard thermal loop resummations in QCD\cite{nair1}. The Chern-Simons
term in this reference has essentially the form
\begin{equation}
L_{CS} \sim \frac{1}{2}\mu(a_+\partial a_- - a_- \partial a_+).
\end{equation}
This is of the general canonical form $p\dot q$. In reference \cite{nair1}
the fictitious gauge field is composed of essentially
the pair $A_\mu$ and $\frac{\delta \Gamma}{\delta A_\mu}$ which are the
natural conjugate variables in the dynamical analogy of the renormalization
group in section 3 if the derivative (dot) is the renormalization scale
rather than the true time. At equilibrium the true time plays no role
as is evident in the imaginary time formalism.
One wonders from the structure in section 3
whether the Chern-Simons term can simply be regarded as the generator
of a renormalization group transformation which induces a gauge
invariant mass for the gauge field. The linear form of
the Lagrangian is strongly suggestive of such an interpretation.
This will be pursued elsewhere.

\section{Non-local sources}

Referring to the equations of motion (\ref{eq:24}-\ref{eq:26}), it is apparent
that
one might extract the field dependence implicit in the sources
by rewriting their expectation values as field operators multiplying new
sources. This would lead to the
treatment of the conjugate problem in place of the finite renormalization
problem\cite{cornwall1}. The contact with renormalization theory
becomes more remote in such a scheme, but the separation of the
problem in terms of field operators is logical in the sense that the need
for extra formal apparatus disappears.

{}From (\ref{eq:24}-\ref{eq:26}), the sources can be formally expanded
in a Taylor series in powers
of the field, leading to a set of relations of the form
\begin{equation}
\frac{\delta^n J_a}{\delta\overline q_{b_1}\cdots\delta\overline
q_{j_n}}\Bigg|_{\overline\phi=0}
= \alpha_1 \delta(x_{b_1},x_{b_2}\cdots,x_{b_n}) + \cdots +
g_n(x_{b_1},x_{b_2}\cdots,x_{b_n})\label{eq:78}
\end{equation}
where the dots indicate all possible mixtures of delta functions with
non-diagonal `matrices' $g_n(x,x',x''\ldots)$. The existence of
non-zero $g$'s follows from the non-locality of the functional
integral. This arises both from its description in terms of
multiple integrals over spacetime quantites and also through its
dependence on the Feynman boundary conditions for propagtors which
depend on the field at widely separated points. For example
\begin{equation}
\frac{\delta J_m}
{\delta\overline\phi(x)\delta\overline\phi(x')}
\Bigg|_{\overline\phi=0}
= \alpha \delta(x,x') + g(x,x').\label{eq:79}
\end{equation}
One observes from the action principle that
the sources can be generated directly from an operator valued
kernel $K$ on replacing $S_{tot}=S+K$, such that in the condensed notation,
\begin{eqnarray}
S_{tot}[\Phi] &=& S[\Phi] + \Lambda + J^a(x)\Phi_a(x)
+\frac{1}{2!}\Phi^a(x) J_{ab}(x,x')\Phi^b(x') \nonumber\\
&+& \frac{1}{3!}J^{abc}(x,x',x'')\Phi_a(x)\Phi_b(x')\Phi_c(x'')\nonumber\\
&+&\frac{1}{4!}J^{abcd}(x,x',x'',x''')\Phi_a(x)\Phi_b(x')\Phi_c(x'')\Phi_d(x''')\nonumber\\
&+& \ldots\label{eq:80}
\end{eqnarray}
and the example considered is of a scalar field theory. Note that the
operator valuedness is only of the quantum type and has no direct connction
with
renormalization or quasi-states.
This effect of this rewriting is to generalize
the notion of sources from linear to non-linear response
in the manner suggested by Dahmen and Jona-Lasinio\cite{dahmen1}, and
computed by Cornwall, Jackiw and Tomboulis\cite{cornwall1}.
Given a particle interpretation, the
sources are $n$-particle sources; they bear no one to one relation to the old
ones.

The effective action for the theory in this form is straightforwardly
generated. Defining
\begin{eqnarray}
\frac{\delta W[J]}{\delta J^a} &=& \overline\phi_a\nonumber\\
\frac{\delta W[J]}{\delta J^{ab}} &=& \frac{1}{2}\lbrack \overline\phi_a
\overline \phi_b + G_{ab}\rbrack\nonumber\\
\frac{\delta W[J]}{\delta J^{abc}} &=& \frac{1}{3!}\lbrack\overline\phi_a
\overline\phi_b\overline\phi_c + 3\overline\phi_{(a}
G_{bc)}+G_{abc}\rbrack\nonumber\\
\frac{\delta W[J]}{\delta J^{abcd}} &=& \frac{1}{4!}\lbrack\overline\phi_a
\overline\phi_b\overline\phi_c\overline\phi_d + 6\overline\phi_{(a}
\overline\phi_b
G_{cd)}+4\overline\phi_{(a}G_{bcd)}+G_{abcd}\rbrack\label{eq:81}
\end{eqnarray}
and performing the Legendre transformation as before,
\begin{eqnarray}
\Gamma[\overline\phi_a,G_{ab},G_{abc},G_{abcd}] &=&
W[J] - J^a\overline\phi_a - \frac{1}{2}J^{ab}\lbrack \overline\phi_a
\overline \phi_b + G_{ab}\rbrack\nonumber\\
&-&\frac{1}{3!}J^{abc}\lbrack\overline\phi_a
\overline\phi_b\overline\phi_c + 3\overline\phi_{(a}
G_{bc)}+G_{abc}\nonumber\rbrack\\
&-&\frac{1}{4!}J^{abcd}\lbrack\overline\phi_a
\overline\phi_b\overline\phi_c\overline\phi_d + 6\overline\phi_{(a}
\overline\phi_b
G_{cd)}+4\overline\phi_{(a}G_{bcd)}+G_{abcd}\rbrack\label{eq:82}
\end{eqnarray}
yields the equations of motion
\begin{eqnarray}
\frac{\delta\Gamma}{\delta\overline\phi^a} &=& -J_a - J_{ab}\overline\phi^b
-\frac{1}{2}J_{abc}(G^{bc}+\overline\phi^b\overline\phi^c)
-\frac{1}{6}J_{abcd}(G^{bcd}+3G^{(bc}\overline\phi^{d)}+\overline\phi^b
\overline\phi^c\overline\phi^d)\nonumber\\
\frac{\delta\Gamma}{\delta G^{ab}} &=& -\frac{1}{2}J_{ab}
-\frac{1}{2}\overline\phi^c
J_{abc}-\frac{1}{4}\overline\phi^c\overline\phi^d J_{abcd}\nonumber\\
\frac{\delta\Gamma}{\delta G^{abc}} &=& -\frac{1}{6}J_{abc}
-\frac{1}{6}\overline\phi^d J_{abcd}\nonumber\\
\frac{\delta\Gamma}{\delta G^{abcd}} &=& -\frac{1}{24}J_{abcd}\label{eq:83}
\end{eqnarray}
and the respective Taylor coefficients
\begin{eqnarray}
\frac{\delta^2\Gamma}{\delta\overline\phi^2}\Bigg|_{\overline\phi=0}
&=& m^2\delta(x_a,x_b) +\frac{\lambda}{2}G_{ab}
- J_{ab} -\frac{1}{2} J_{abcd}G^{cd}\nonumber\\
\frac{\delta^3\Gamma}{\delta\overline\phi^3}\Bigg|_{\overline\phi=0}&=&
-J_{abc}\nonumber\\
\frac{\delta^4\Gamma}{\delta\overline\phi^4}\Bigg|_{\overline\phi=0}
&=& \lambda \delta(x_a,x_b,x_c,x_d)+\frac{3\lambda^2}{2}G_{abcd} -
J_{abcd}.\label{eq:84}
\end{eqnarray}
The latter relations imply that
\begin{eqnarray}
\frac{\delta^2\Gamma}{\delta\overline\phi^2}\Bigg|_{\overline\phi=0}
&=& m^2\delta(x_a,x_b) +\frac{\lambda}{2}G_{ab}+
2 \frac{\delta\Gamma}{\delta
G_{ab}}-\frac{1}{2}G^{cd}\frac{\delta^4\Gamma}{\delta\overline\phi^4}\Bigg|_{\overline\phi=0}\nonumber\\
\frac{\delta^3\Gamma}{\delta\overline\phi^3}\Bigg|_{\overline\phi=0}
&=& 3 \frac{\delta\Gamma}{\delta G_{abc}}\nonumber\\
\frac{\delta^4\Gamma}{\delta\overline\phi^4}\Bigg|_{\overline\phi=0}
&=& \lambda \delta(x_a,x_b,x_c,x_d)+\frac{3\lambda^2}{2}G_{abcd}+
4 \frac{\delta\Gamma}{\delta G_{abcd}}.\label{eq:85}
\end{eqnarray}
which illustrates nicely the triality between $G^{(n)}$,
$\frac{\partial^n\Gamma}{\partial\overline\phi^n}$ and
$\frac{\delta\Gamma}{\delta G^{(n)}}$.
The new source terms $J^{(n)}\equiv J^{a_1\ldots a_n}$
generate non-local counterterms  $R^{ij}_\alpha\xi^\alpha$
and the preceding methodology is apparent in a modified form.
See (\ref{eq:47}) and (\ref{eq:110}).
In particular,
if one considers the action principle for variations with respect
to some common generator $\xi^\alpha = \mu^2$, one has
\begin{equation}
\delta\langle 0 \infty |0 -\infty \rangle =\langle 0 \infty |\delta\Big(
S + \Lambda + J^a\Phi_a + \frac{1}{2}\Phi^a J_{ab}\Phi^b+\ldots
\Big)|0 -\infty\rangle\label{eq:86}
\end{equation}
which, using the equations of motion (\ref{eq:83}), should lead to the
corresponding
renormalization group equation in this new scheme
\begin{eqnarray}
\Bigg[
\frac{\partial}{\partial\mu}
-\beta^{(n)}_i\;\frac{\delta}{\delta G^{(n)}_i}
\Bigg]\Gamma = 0.\label{eq:87}
\end{eqnarray}
Where $\beta^{(n)}=\frac{\partial G^{(n)}}{\partial\mu}$,
$G^{(1)}=\overline\phi$.
This result is what one would naively expect
on differentiating a function of $N$ $n$-point functions.
On performing the variation however, one obtains
\begin{eqnarray}
\Bigg[
\frac{\partial}{\partial\mu}
-\beta_\phi^a \frac{\delta}{\delta\overline\phi^a}
-\beta_{G,ab}\frac{\delta}{\delta G_{ab}}
-\beta_{G,abc}\frac{\delta}{\delta G_{abc}}
-\beta_{G,abcd}\frac{\delta}{\delta G_{abcd}} + A(\overline\phi^a,G^{ab}\ldots)
\Bigg]\Gamma = 0\label{eq:88}
\end{eqnarray}
where,
\begin{equation}
A = 4\beta_\phi^a G^{bcd}\frac{\delta \Gamma}{\delta G^{abcd}}
+\beta_{G}^{ab}\Big( -\frac{\delta \Gamma}{\delta G^{ab}}
+6\overline\phi^c \frac{\delta\Gamma}{\delta G^{abc}}
-13\overline\phi^c\overline\phi^d\frac{\delta\Gamma}{\delta G^{abcd}}
 \Big)\label{eq:90}
\end{equation}
The anomalous term $A$ arises due to the truncation of the series of
sources at some level.
The interpretation of the result is as follows. When $G^{(n)}\equiv
G^{(i_1\ldots i_n)}$ is the exact solution to the equation
$\frac{\delta\Gamma}{\delta G^{(n)}}=0$, the invariance of the
transformation function follows automatically from the vanishing of the
sources in (\ref{eq:83}). Suppose now that $G^{(n)}$ is not an exact solution;
this means that the source in non-zero, implying in turn that
an effective non-zero counterterm is in the theory
in order to satisfy (\ref{eq:87}) exactly. When the result is
exact there is clearly no computational advantage in
adding finite counterterms. The normal
situation is that all of the $G^{(n)}$ are approximations, in which case
one obtains a renormalization group type equation relating equivalent
theories to vertex rescalings. It is worth noting that the truncation
of this series at quadratic order corresponds to a self-consistently
defined mass counterterm, but no quartic coupling counterterms.
While this is clearly the most important contribution for inhibiting
infra-red divergences, it does not take into account the fact that
the other couplings in the theory are also effectively altered by
higher order corrections. This deficiency will be most apparent in
models where the ratio of two couplings
determines some crucial physical property; the phase
transition in a gauge theory is an example of this.

One notes briefly that to extend the discussion to encompass
Abelian gauge theories, it is necessary to consider
mixed sources in an arbitrary gauge. The discussion in section 5
indicates that the sources must take the general form
\begin{eqnarray}
S_{source} &=& \int dV_x dV_{x'} \big\lbrace
\frac{1}{2}\Phi^\dagger(x) \Delta(x,x')\Phi(x')
+\Phi(x) C^\mu(x,x') A_\mu \nonumber\\
&+& \Phi^\dagger(x) C^{\mu\dagger}(x,x') A_\mu(x')
+\frac{1}{2}A^\mu(x) J_{\mu\nu} A^\nu(x').
\big\rbrace\label{eq:91}
\end{eqnarray}
Under a gauge transformation $\Phi\rightarrow e^{i\theta(x)}\Phi(x)$,
$A_\mu \rightarrow A_\mu(x)+\partial_\mu \theta(x)$, one finds that
\begin{eqnarray}
\frac{\delta S}{\delta \theta(x)} &=& \int dV_x \big\lbrace
iC^\mu(x,x')(A_\mu(x')+\partial_\mu\theta(x'))
-iC^{\mu\dagger}(x,x')(A_\mu(x')+\partial_\mu\theta(x'))\nonumber\\
&-& i\theta(x')\stackrel{x}{\partial_\mu}C^\mu(x,x')
+ i\theta(x')\stackrel{x}{\partial_\mu}C^{\dagger\mu}(x,x')\nonumber\\
&-& 2\stackrel{x}{\partial_\mu}J^{\mu\nu}(x,x')A_\nu(x')
+2\stackrel{x}{\partial_\mu}\stackrel{x'}{\partial_\nu}
J^{\mu\nu}(x,x')\theta(x')
\big\rbrace\label{eq:92}
\end{eqnarray}
and thus gauge invariance requires
\begin{eqnarray}
C^{\dagger\mu}(x,x') &=& C^{\mu}(x,x')\nonumber\\
\stackrel{x}{\partial_\mu}J^{\mu\nu}(x,x') &=& 0\label{eq:93}
\end{eqnarray}
where all the objects are symmetrical in $x,x'$. While $C^\mu(x,x')=0$
is a solution to the gauge invariance condition, it is clear from
the preceding section that the $C^\mu$'s play an important role
in maintaining the structure of the gauge coupling.

A note is in order concerning the relationship between the renormalization
group and the present method. Since the renormalization group can only
reorganize information is already in the loop expansion, it must be completely
equivalent to a resummation in the final analysis. Differences
are nonetheless apparent. The renormalization group focusses on
the couplings of a theory in a democratic way, while most resummation schemes
favour effective masses for practical reasons. On the other hand, the
beta-functions used in
the RG analysis are normally derived from low order results and thus
the validity of conclusions arrived at in the RG is limited by
the extent to which the perturbation series can be trusted. Once does
not escape the bounds of self-consistency.
Both methods resum only particular terms which fit their formal
structure -- there is no guarantee that the terms excluded from such
privileged resummations are not important, so this must be argued
independently.
The method
described in section 4 has the distinct advantage of demonstrating the
formal relationship between the two methods, but as long as
non-perturbative calculation demands a certain cunning, it will be
important to cross check the different approaches.

\section{Open dynamical systems and non-equilibrium}

In a many particle theory one is naturally led to consider the effect of
disturbances which drive the field into a state of disequilibrium\footnote{For
recent work with reference to various
approaches, see ref \cite{hu2,heinz1,eboli1}.}. Such a
disturbance may lead to one of two formally distinct situations:
(i) homogeneous, time-dependent energy distributions, (ii)
inhomogenous, time-dependent mass-energy distributions. In a
closed system one may only have the latter type, since energy
cannot escape, it can only be redistributed about the system.
In the absence of external forces or sources a system returns,
by means of collision and random motion to a homogeneous,
time-independent state within some characteristic relaxation
time $\tau$. If the initial state is inhomogenous, this decay
process involves transport.

In recent work (see for example \cite{calzetta1,lawrie1,eboli1})
time dependent coupling constants have been used to model non-equilibrium
systems. It is interesting in the light of the dynamical analogy
presented in this paper to ask whether such a time variation of
couplings might simply be regarded as a renormalization group flow
enacted in real time. This has been tacitly assumed by some authors.
To answer this question it is important to understand that a system
with time-dependent couplings represents at best
a phenomenological description of a physical system in the same
way that renormalized coupling constants are phenomenological in the
sense that they simply
absorb the effect of virtual interactions into `effective' parameters.
Thus one is interested in knowing: to what extent is a theory with
time dependent coupling constants simply a time dependent
rescaling of a certain renormalizable theory?

The answer to this question
depends on the extent to which the time variation of the coupling
constants can be thought of as describing a symmetry of the system.
The answer is evidently that, if the values of the couplings are
governed by a Hamiltonian, then the answer to the above question
must be yes. However, this is dependent on the way in which the
notion of time-dependent couplings is used.
One could imagine simply specifying that all the couplings should
increase linearly with time -- but this would not be a local physical
system in the normal sense.
In systems possessing a conserved
current one cannot {\em a priori} expect to simply substitute time
dependent couplings into the action with impunity. General invariance
under symmetry transformations will dictate the class of variations of
the action for which the field equations and
evolution generators preserve those symmetries and
can therefore be considered physical\cite{burgess6}.

To summarize: in a closed system one is
guided by principles of energy conservation, gauge invariance and
other conservation laws, but in an open system any one of these
bastions of principle can be locally violated.
Indeed it is often stated that Lagrangian or Hamiltionian descriptions
of open systems do not exist, though this is an exaggeration since
certain open systems can in fact be regarded as constrained
systems\cite{burgess6} if there is sufficient symmetry on which to
intuit legal behaviour.
The present discussion will be restricted to those limited cases in which
a Lagrangian methodology is appropriate.

Consider the closed total system,
\begin{equation}
S_{tot}=S_1(x_1)+S_{12}(x_1,x_2)+S_2(x_2)\label{eq:94}
\end{equation}
composed of two subsystems which interact through the contact term
$S_{12}$. $S_{12}$ serves as a source both supplying perturbations
from $S_2$ to $S_1$ and the backreaction of $S_1$ on $S_2$. In thermal
field theory, $S_2$ is often identified with a heat bath which is
sufficiently large that this backreaction problem can be
neglected in a first approximation. Since $S_{tot}$ is closed
one may write
\begin{equation}
\frac{d \langle H_{tot}\rangle}{dt} = 0, ~~~\frac{\delta S_{tot}}{\delta
\theta_i(x)} = 0 ~~~\ldots\label{eq:95}
\end{equation}
for symmetry generators $\theta_i(x)$ and $x$ covers the whole system.
However, for the subsystems in general\footnote{This lesson
applies also to an isolated unitary field theory. If one truncates
part of the system by working to some order in perturbation
theory, the perturbative result will behave like a partial system
and will exhibit apparent dissipative characteristics. The remainder
of the uncalculated terms behave as the external part. The
issue of gauge fixing dependence is also known to be related
in some cases to the
truncation of the generating functional at some perturbative order.
The question of sources and conservation of
degrees of freedom is also connected with ghosts and
unitarity\cite{feynman1}.},
\begin{eqnarray}
\frac{d\langle H_a\rangle}{dt} \not=&0&\nonumber\\
\frac{\delta S_a}{\delta\theta_i(x)} \not=&0&.\label{eq:96}
\end{eqnarray}
In the above notation
this involves a restriction to $S_1$. The effect of $S_2$ on the system is
then modelled by the time variation of the coupling constants in
$S_1$. This bears of course the additional assumption that such a division
leads to a phenomenological description in terms of these couplings.
This point conceals an important and fundamental
ambiguity which will be discussed below.

The physical conservation laws of the reduced system are
not those of the total system, but they must lead internally to
a well defined set of observables in $S_1$. It is noteworthy that,
if one assumes the existence of time dependent couplings
without explicitly introducing a physical source which generates them,
then they may also have to be spatially dependent in order not to
violate conservation laws\cite{burgess6}.
The analogy with polarons with interaction-induced position
dependent masses is evident; in this case the interaction
with the local system acts as a source. If
one insists on purely time dependent couplings then conservation laws
can only be upheld if they represent true fields in an extended
Hamiltonian, or if one explicitly introduces external sources which
generate the change.

An elementary example in which a difficulty arises is in a leaky
gauge theory such as might be obtained by connecting a closed
system to a battery (an external system) using leads (a contact term).
Consider the minimal coupling of the gauge field to
a current $J_\mu$,
\begin{equation}
S = \int dV_x\; J^\mu A_\mu.\label{eq:97}
\end{equation}
Let the integral over $x$ contain some boundary $\Sigma$ which is
permaeable to $J^\mu$. The system is open, but the obervable coordinates
$x_1$ are restricted to the interior of the boundary. Under a gauge
transformation $A_\mu\rightarrow A_\mu+\partial_\mu\theta$,
\begin{equation}
\delta S_1 = \int dV_{x_1}\lbrace -(\partial^\mu J_\mu)\theta
+\partial^\mu (J_\mu\theta)\rbrace.\label{eq:98}
\end{equation}
Since the current is not conserved within the domain of $x_1$, this
quantity is non-vanishing in the vicinity of the boundary $\Sigma$.
It should be clear that the source of the deficiency is in the
construction of the theory and not in its calculation. The only
solution is to postulate the remainder of the system (in this case the
battery) on the basis
of conservation arguments or to include a dissipative `integrating
factor' in the Lagrangian, which fixes the formal consistency at
the expense of the introduction of new variables\cite{burgess2,burgess6}.
If one insists on describing a partial system with time dependent couplings,
difficulties in the construction of non-equilibrium
systems cannot be entirely eschewed: they involve a fundamental
ambiguity in the manner in which the partitioning of the
subsystems is achieved.

The action principle for the direct computation of statistical
expectation values at equilibrium or otherwise, has been
given by Schwinger\cite{schwinger2} and leads to quantities of the
form
\begin{equation}
\langle t_2|{\cal O}(t_1)|t_2\rangle\label{eq:99}
\end{equation}
where $|t_2\rangle$ describes some macrostate of the system. Since
the time-dependence of the operator is now an issue, it matters that the
bra and ket states be described by the same basis. (\ref{eq:99}) then gives the
expectation
value of the operator $\cal O$ at time $t_1\rangle t_2$ given that the system
was in the prescribed state $|t_2\rangle$ at $t_2$. Now
\begin{equation}
\langle t_2|{\cal O}(t_1)|t_2\rangle = \sum_{i,i'} \langle t_2|i\rangle
\langle i|{\cal O}(t_1)|i'\rangle\langle i'|t_2\rangle\label{eq:100}
\end{equation}
which involves the mutually conjugate forms of the action principle
\begin{eqnarray}
\delta \langle t_2|i\rangle&=& i \langle t_2|\delta S|i\rangle\nonumber\\
\delta \langle i|t_2\rangle &=& -i \langle i|\delta S|t_2\rangle.\label{eq:101}
\end{eqnarray}
Since the time-dependence of an operator advances by the unitary rule
$U\,{\cal O} U^{\dagger}$, for which one may solve
$U(t_2,t_1)=T\exp(-i\int_{\Sigma_1}^{\Sigma_2}J'\Phi)$,
$U^{\dagger}=T^{\dagger}\exp(i\int_{\Sigma_1}^{\Sigma_2}J'\Phi)$
for spacelike hypersurfaces $\Sigma_a$, the
expectation value with respect to equal in-out states involves both
time-ordered and anti-time-ordered transformation amplitudes. Both
forms are available through (\ref{eq:101}) and the complete expression can be
written down without reference to the individual amplitudes or their
conjugates:
\begin{equation}
\delta \langle t_2|t_2\rangle =
i \langle |\int_{\Sigma_1}^{\Sigma_2}dt({\cal L_+ -
L_-})|\rangle.\label{eq:102}
\end{equation}
The distinction between ${\cal L_+}$ and ${\cal L_-}$ is for formal
convenience, since, on introducing the sources $J_+$ and $J_-$
for the unitary and anti-unitary transformations, one may write the
generating functional for the appropriate combinations of Green
functions
\begin{equation}
\delta \langle t_2|t_2\rangle =
i \langle |\delta(S_+[\Phi_+]+J_+\Phi_+ - S_-[\Phi_-] -J_-\Phi_-)|\rangle,
\label{eq:103}\end{equation}
so that
\begin{equation}
-i\frac{\delta}{\delta J_+(t_1)} \langle t_2|t_2\rangle \Bigg|_{+=-}
= i\frac{\delta}{\delta J_-(t_1)} \langle t_2|t_2\rangle\Bigg|_{+=-}
=\langle t_2|\Phi(t_1)|t_2 \rangle.\label{eq:104}
\end{equation}
The result generalizes for other expectation values.

Whilst there is no ambiguity in the usage of ref. \cite{schwinger2},
issues of definition inevitably haunt the fringes of
the calculational procedure for open systems.
The situation of interest here is that in
which the phenomenological couplings are time- or spacetime-dependent.
Appealing to the discussion in section 4, one understands that
the couplings are themselves then fields. Indeed, this is no longer
a device but a phenomenological `reality'. In particular one must understand
how to define $\langle t_2|t_1\rangle$ when, in passing from $t_1$
to $t_2$ or vice-versa, the values of the couplings have changed.
While in certain adiabatic circumstances it might be possible to
neglect this change, in general the elevation of this remark to
a precise statement is the source of the fundamental ambiguity in the
construction of a phenomenological model of an open system.
In the partitioning of a system into subsystem $S_1$
and 'reservoir' $S_2$ one is led to consider the effect of $S_1$ on
$S_2$, but often neglecting the effect that such change has on
the $S_1$ -- the so-called back-reaction problem. Often one is interested
in the case where $S_1$ is a large reservoir so that the back-reaction
is negligeable, nevertheless it is this neglect of the back-reaction which
gives the illusion of dissipation. To take account of this backreaction in
a phenomenogical system, it is necessary to complete the logical system
in some way. Since in general
one does not possess any detailed knowledge about such a reservoir,
it is necessary to do this in a fairly arbitrary manner. Consider the
action
\begin{equation}
S_s = \int \;dV_x\Bigg\lbrace \frac{1}{2}(\partial^\mu\phi)(\partial_\mu\phi)
+\frac{1}{2}m^2 \phi^2+\xi R(x)\phi^2 +\frac{\lambda}{4!}\phi^4 \Bigg\rbrace
\label{eq:a1}
\end{equation}
where $R(x)$ might be regarded as the extrinsic curvature of some lower
dimensional
subsystem embedded in a higher-dimensional
space. (This could pertain to excitations trapped
in a boundary layer, for instance\cite{burgess7}.) Let $R(x)$
satsify the phenomenological equation
\begin{equation}
(-\boxx+\omega_0^2) R(x)=J_{R\phi}\label{eq:a2}.
\end{equation}
If the source $J_{R\phi}$ is vanishing in (\ref{eq:a2}), an asymmetry is
introduced into the formulation. From (\ref{eq:a1}) it is evident that
$R(x)$ is a source for $\phi^2$, however the converse is only true if
$J_{R\phi}\not=0$ in (\ref{eq:a2}). The latter corresponds to neglecting
the back-reaction of the system defining $R(x)$. It is clearly
incorrect to neglect this change in a physical system: even though it
may be small in real terms, it has an important formal function.
Internal consistency requires that $R(x)$ satisfy an equation of motion
which is compatible with the remainder of the system. It must be
``on-shell''.
A more
realistic approach is to rewrite (\ref{eq:a1}) and (\ref{eq:a2})
to give the local form
\begin{equation}
S\rightarrow S+\int dV_x\Bigg\lbrace\alpha R(-\boxx+\omega_0^2)R\Bigg\rbrace
\end{equation}
where $\alpha/\xi$ measures the relative strength of the contact. The
implication is that a time dependent
effective mass is not a complete description in itself, but a better
simulation of a physical system
is obtainable on including the kinematical development through
a source coupling as in a local theory. Clearly this theory is different
from the one in which one ignores changes dependent changes in $R(x)$,
but it shares more of the characteristics of a fundamental system than
the alternative. As emphasized by Schwinger, sources play an important
conceptual role. A source may either be fictitious, as a generator of
some transformation, or physical in the manner of an external
generalized force. In an open system, this distinction is blurred.

Given a phenomenological open system, one must proceed to
calculate it. Infrared behaviour must be regularized in an
analoguous way to that in section 4.
Renormalized couplings must now be
defined at a given energy and at a given time (this distinction is
no longer clear).
The previous methods in section 4 can be imported. Sources for the couplings
have an exactly equivalent role and the action principle takes
on the form
\begin{equation}
\delta \langle t_2|t_2 \rangle= i\langle t_2|\delta \tilde{S}| t_2\rangle
\label{eq:105}\end{equation}
where
\begin{equation}
\tilde{S}=\int_{\Sigma_1}^{\Sigma_2}\lbrace \frac{1}{2}q^i D_{ij}q^j
+\ldots \rbrace\label{eq:106}
\end{equation}
and $q^i = \lbrace \Phi_+,\Phi_-,m^2_+,m^2_-,\ldots\rbrace$. $D_{ij}$
is a generalized derivative operator, which encodes the appropriate
signs in (\ref{eq:103}). For the couplings, this must be determined from their
equations of motion. If not all the $q^i$ are true degrees of
freedom in the reduced system, there might be subsidiary conditions
to satisfy; in the above covariant form the problem is then reminiscent
of a gauge theory. The counterterms in (\ref{eq:47}) etc. are now
matrix valued and may contain terms relating to dissipation if one
uses a model in which back-reaction is neglected. These two will
contribute to screening infra-red divergences.
The method of Lawrie\cite{lawrie1} appears to coincide
with that described here but neglects
the backreaction problem discussed above;
it provides an example of the computation of
a mass counterterm when correlations between the time dependent masses
are not coupled to the subsystem minimally, that is, when the subsequent
development of the system has no effect on the effective mass itself.
A further pleasing demonstration can be found in ref \cite{calzetta2}.
The calculations
of this paper will not be repeated here, but it is useful to discuss
the result from the present perspective. The authors consider
particle production (vacuum dissipation) due to an isolated single-particle
source. Starting with the action (in the original conventions)
\begin{equation}
S=\int dV_x \Bigg\lbrace \frac{1}{2}(\partial^\mu\phi)(\partial_\mu\phi)
-\frac{1}{2}m^2\phi^2-\frac{1}{6}g\phi^3 -J\phi\label{eq:b1}
\Bigg\rbrace,
\end{equation}
the closed time path effective action can be written
\begin{equation}
\Gamma[\overline\phi_+,\overline\phi_-] = -i\ln\int D\varphi_+ D\varphi_-
\exp(iS[\overline\phi_++\varphi_+]-iS[\overline\phi_-+\varphi_-]-\varphi_+\Gamma[\overline\phi_+]_{,+}+-\varphi_-\Gamma[\overline\phi_-]_{,-}).
\end{equation}
Whereas the usual effective action can often be interpreted literally
as an effective action, the present object has no immediate interpretation.
Only its variation with respect to one of the arguments
can be interpreted after setting the $+$ fields and sources equal
to the $-$ ones. From (\ref{eq:28}) one has effective field equations
for the physical expectation value $\langle t_2|\phi(t_1)|t_2\rangle$
\begin{equation}
\Gamma[\overline\phi_+,\overline\phi_-]_{,i_+} \Bigg|_{+ = -} = \langle
t_2|S_{,i}(t_1)|t_2\rangle,
\end{equation}
if the functional differentiation is with respect to $\overline\phi_+(t_1)$.
The linearized field equations to order $g^2$ were deduced to be of the
form
\begin{equation}
(\boxx+m^2 +g^2 Z)\overline\phi(x) = J(x)
\end{equation}
for some complex operator $Z$. The source term in this expression plays
an analogous role to the time-varying mass in the previous example.
It acts as a source for the expectation value of the field. Since no
source-source correlations are included, there is no backreaction
on the source and an imaginary part is developed due to the non-conservation
of the incomplete problem.
This is, by the discussion in previous sections, equivalent to the
claim by the authors that higher non-local sources will rectify
the source of the dissipation. Some examples will be presented in a
separate publication.

A final comment is in order concerning the infrared problem in non-equilibrium
systems. In the present formalism, the cure proceeds by direct analogy
with the previous discussion, prior to taking the limit $+ \rightarrow -$
in the auxiliary fields. The essential difference is the matrix valued
nature of the propagators. All of the previous limitations
and cautions apply.
More explicit calculations for open systems will be deferred until a later
publication.

\section{Concluding remarks}

The action principle, as given by Schwinger, is an elegant way of obtaining
amplitudes and expectation values in field theory. Emphasis is
placed on the role of unitary or canonical transformations. By introducing
sources for the couplings, renormalization group transformations
can be written in a form which is obtainable from the action principle.
The structure one generates in this way enables familiar results from
renormalization group analysis to be obtained through self-consistent
equations for the propagators. This is strongly reminscent of the method
of non-local sources and can be related to it by a reparameterization.

The present formalism shows that phenomenological models with time-dependent
couplings can be thought of as realtime enactions of renormalization
group transformations only if the time variation is in accordance
with a generalized Hamiltonian for both the fields and couplings.
In other instances such models must be regarded as open systems and one cannot
expect
them to respect conservation laws or their associated symmetries.

As remarked in section 4 and demonstrated in reference \cite{nair1},
the Chern-Simons
Lagrangian has a symplectic form which leads to the thermal
mass resummation in the the high temperature infrared regime of
QCD. It is interesting to speculate as to whether a more
general connection with infrared behaviour can be established for gauge
theories. The present dynamical analogy suggests that the Chern-Simons
action is simply the generator of a renormalization group
transformation, but a more careful
investigation is needed to formalize a connection. This and other
issues will be considered in subsequent work.

\section*{Acknowledgement}
I would like to thank Gabor Kunstatter for an invitation to the University of
Winnipeg where part of the present work was done.
I am specially grateful also to Ian Lawrie for patient discussions about
the traditional renormalization group in field theory and for his
helpful criticisms of this work. I am grateful to Julian Schwinger
for rekindling my interest in sources.
Finally, I would like to thank Jens Andersen for showing me his result for
the 2-loop calculation in appendix A and also
Meg Carrington, Bei-Lok Hu and David Toms and C. Stephens
for their comments on the revised manuscript.
\bigskip

This paper is dedicated warmly to the memory of Julian Schwinger.

\appendix
\section{Regularized $\lambda\phi^4$ interaction}
Consider the Euclideanized action
\begin{eqnarray}
S = \int dV_x \lbrace \frac{1}{2} (\partial_\mu\varphi)(\partial^\mu\varphi)
+\frac{1}{2}m^2\varphi^2+\frac{\lambda}{4!}\varphi^4\rbrace
\label{eq:107}\end{eqnarray}
To one loop order, the effective potential is given by
\begin{eqnarray}
V_{eff}= \frac{1}{2}m^2\overline\phi^2 + \frac{\lambda}{4!}\overline\phi^4
+\frac{(m^2+\frac{\lambda}{2}\overline\phi^2)^2}{64\pi^2}\lbrack
\ln\left(\frac{m^2 + \frac{\lambda}{2}\overline\phi^2}{\mu^2}\right)
-\frac{\alpha}{2}\rbrack\label{eq:108}
\end{eqnarray}
where $\alpha$ is a divergent constant, the details of which vary for
different regularization schemes. The following (re)normalization
conditions are imposed:
\begin{eqnarray}
V(0) &=& 0\label{eq:109}\\
\frac{\partial^2 V_{eff}}{\partial\overline\phi^2}\Bigg|_{\overline\phi=\Phi_m}
&=& m^2 + \frac{\lambda}{2}\Phi^2_m \equiv M^2\label{eq:110}\\
\frac{\partial^4
V_{eff}}{\partial\overline\phi^4}\Bigg|_{\overline\phi=\Phi_\lambda}
&=& \lambda\label{eq:111}
\end{eqnarray}
Let $m^2\rightarrow\overline m^2 +\delta m^2$ and $\lambda\rightarrow
\overline\lambda+\delta\lambda$. The above conditions now fix the counterterms.
Specifically, to leading order, one finds
\begin{eqnarray}
\delta \lambda = \frac{3\overline\lambda^2}{32\pi^2}\Bigg\lbrack
\ln\left(\frac{\overline m^2 +
\frac{\overline\lambda}{2}\overline\phi^2}{\mu^2}\right)
-\frac{\alpha}{2}\Bigg\rbrack + \frac{9\overline\lambda^2}{64\pi^2}
-\frac{\overline\lambda^4\Phi_\lambda^4}{32\pi^2(\overline
m^2+\frac{\overline\lambda}{2}\Phi^2_\lambda)^2}
+\frac{3\overline\lambda^3\Phi^2_\lambda}{16\pi^2(m^2+\frac{\overline\lambda}{2}\Phi^2_\lambda)}\label{eq:112}
\end{eqnarray}
\begin{eqnarray}
\delta m^2 &=& \lbrack -\frac{\overline\lambda\Phi^2_m}{32\pi^2}
-\frac{\overline\lambda M^2}{32\pi^2}
+\frac{3\overline\lambda^2\Phi^2_m}{64\pi^2} \rbrack\Bigg \lbrack
\ln\left(\frac{\overline m^2 +
\frac{\overline\lambda}{2}\overline\phi^2}{\mu^2}\right)
 -\frac{\alpha}{2}\Bigg\rbrack
-\frac{3\overline\lambda^2\Phi^2_m}{64\pi^2}\nonumber\\
 &-&\frac{\overline\lambda M}{64\pi^2}
+\frac{9\overline\lambda^2\Phi_m^2}{128\pi^2}
- \frac{\overline\lambda^4\Phi^4_\lambda\Phi^2_m}{64\pi^2(\overline
m^2+\frac{\overline\lambda}{2}\Phi^2_\lambda)^2}.\label{eq:113}
\end{eqnarray}
For the investigation of critical phenomena one is interested in the massless
limit $\overline m^2=0$,
whereupon the effective potential becomes
\begin{eqnarray}
V_R(\overline\Phi) = \frac{\overline\lambda}{4!}\overline\phi^4 +
\frac{9\overline\lambda\Phi^2_m\overline\phi^2}{128\pi^2}
+\frac{3\overline\lambda^2\Phi^2_m\overline\phi^2}{128\pi^2}\ln\Bigg(\frac{\Phi^2_\lambda}{\Phi^2_m} \Bigg)+\frac{\overline\lambda^2\overline\phi^4}{256\pi^2}
\Bigg\lbrack \ln\Bigg(\frac{\overline\phi^2}{\Phi_\lambda^2}\Bigg) -
\frac{25}{6}\Bigg\rbrack\label{eq:114}
\end{eqnarray}
There is no particular loss from taking $\Phi_m=\Phi_\lambda\equiv\Phi$ and one
sees
that, in the limit that $\overline\phi\rightarrow 0$ one must regularize the
logarithms by taking $\Phi\sim\overline\phi$. This has the effect of
resumming the logs to leading order and results in a potential which
goes purely like $\overline\phi^4$, a result which may be shown from
the renormalization group. Clearly the larger $\Phi^2$ becomes,
the less reliable the original Taylor series expansion becomes and thus
one cannot draw conclusions about the large $\overline\phi$ region with
any certainty. The present result only shows that there is no
discontinuous change in the potential (the potential
rises only positively) as the minimum moves away from
the origin, implying a continuous, second order phase transition.

In $2+1$ dimensions the unrenormalized potential is given by
\begin{equation}
V_{eff}(\overline\phi) = \frac{1}{2}m^2\overline\phi^2
+\frac{\lambda}{4!}\overline\phi^4 -
\frac{(m^2+\frac{\lambda}{2}\overline\phi^2)^{\frac{3}{2}}}{12\pi}\label{eq:115}
\end{equation}
to one loop order.
Logarithmic corrections to this order are conspicuous by their absence;
an arbitrary scale now lies hidden in the dimensionful couplings.
This follows in turn from the absence of ultraviolet divergences\footnote{This
result is of course only a convenient fiction, since the statement is dependent
on the way in which one chooses to `regulate' the divergences. There are
divergences which we choose not to consider since they do not enter in
ratios with other scales and cannot therefore be significant to the
underlying physics.} although at two loops there is a divergent
contribution. To show this, we note that
\begin{equation}
\Gamma^{(2)} = \langle
S^{(4)}\rangle-\frac{1}{2}\langle(S^{(3)})^2\rangle\label{eq:116}
\end{equation}
where $S^{(n)}$ is the part of the classical action of $n$-th order
in the quantum fields after making the background field split
$\varphi\rightarrow\varphi+\overline\phi$. The operator product average
may be defined by
\begin{equation}
\langle F[\varphi]\rangle = \frac{\int d\mu[\varphi] F[\varphi]
e^{-\frac{1}{h}S^{(2)}}}{\int d\mu[\varphi]
e^{-\frac{1}{h}S^{(2)}}},\label{eq:117}
\end{equation}
and only the one-particle irreducible terms survive. Using Wick's theorem
it is straightforward to show that
\begin{eqnarray}
\langle S^{(4)}\rangle &=& \int dV_x
\frac{\lambda}{8}\Delta(x,x)\Delta(x,x)\label{eq:118}\\
\langle(S^{(3)})^2\rangle &=& \int dV_x
dV_{x'}\frac{\lambda^2\overline\phi^2}{6}
\Delta(x,x')\Delta(x,x')\Delta(x,x')\label{eq:119}
\end{eqnarray}
The first `graph' is finite after regularization
\begin{eqnarray}
\Delta(x,x) &=&
\int\frac{d^3k}{(2\pi)^3}\frac{1}{k^2+m^2+\frac{\lambda}{2}\overline\phi^2}\nonumber\\
&=& - \frac{|m^2+\frac{\lambda}{2}\overline\phi^2|}{4\pi},\label{eq:120}
\end{eqnarray}
whereas the latter diverges:
\begin{eqnarray}
I &=&\int dV_x dV_{x'} \Delta(x,x')\Delta(x,x')\Delta(x,x')\nonumber\\
&=& \int dV_x \int \frac{d^nk_1}{(2\pi)^n}\frac{d^nk_2}{(2\pi)^n}
\frac{\mu^{-2\epsilon}}{(k_1^2+{\cal M}^2)(k_2^2+{\cal M}^2)((k_1+k_2)^2+{\cal
M}^2)}\label{eq:121}
\end{eqnarray}
Following Collins, the denominators are combined using
\begin{equation}
\frac{1}{a_1a_2a_3}=2\int_0^1
d^3z\delta(1-z_1-z_2-z_3)[a_1z_1+a_2z_2+a_3z_3]^{-3}\label{eq:122}
\end{equation}
whereupon
\begin{eqnarray}
I &=& (4\pi)^{-n}\Gamma(3-n)\int dV_x ({\cal M}^2)^{n-3} G(n)\mu^{-2\epsilon}
\label{eq:123}\\
G(n)&=& \int_0^1
d^3z\;\delta(1-z_1-z_2-z_3)(z_1z_2+z_2z_3+z_3z_1)^{-\frac{n}{2}}.\label{eq:124}
\end{eqnarray}
$G(3)$ is evaluated on making the subsitution $z_1=\rho x$, $z_2=\rho(1-x)$,
$z_3=1-\rho$, which implements the delta-function constraint
explicitly, giving
\begin{equation}
G(3)=\int_0^1\frac{\rho d\rho
dx}{(\rho-\rho(1-x+x^2))^{\frac{3}{2}}}\label{eq:125}
\end{equation}
Note that although the integrand is singular, the
divergence is illusory and the integral itself
is finite. Evaluating this integral leads to
\begin{equation}
G(3) = 2\int_0^1\frac{dx}{\sqrt(x-x^2)} = 2\pi\label{eq:126}
\end{equation}
Thus, one obtains
\begin{equation}
I = -\frac{1}{32\pi^2}\Bigg\lbrack
\frac{1}{\epsilon}+\ln\Bigg\lbrack\frac{{\cal M}^2}{4\pi\mu^2}\Bigg\rbrack +
\gamma\Bigg\rbrack\label{eq:127}
\end{equation}
where $n=3+\epsilon$, ${\cal M}^2=m^2+\frac{\lambda}{2}\overline\phi^2$ and
$\gamma$ is Euler's constant. This shows that there is a single logarithm
of the effective mass at two loop order.

Considering only the one-loop potential (nonetheless mindful of the
two-loop logarithm), one proceeds to apply the renormalization
conditions (\ref{eq:109}-\ref{eq:111}) which leads, in the massless limit, to
\begin{eqnarray}
\delta\lambda &=& 0\label{eq:128}\\
\delta m^2 &=& \frac{\overline\lambda^{3/2}\Phi_m}{4\pi\surd 2}\label{eq:129}\\
V_R(\overline\phi) &=& \frac{\overline\lambda\overline\phi^4}{4!}
+\frac{\overline\lambda^{3/2}\Phi_m}{8\pi\surd 2}\overline\phi^2
-\frac{\overline\lambda^{\frac{3}{2}}\overline\phi^3}{24\surd 2
\pi}.\label{eq:130}
\end{eqnarray}
There is a subtlety here, namely that $\lambda$ is not a dimensionless
quantity, and so it is not possible to disuss its smallness directly.
To this end one may introduce $\lambda = \Phi_s^2\overline\lambda^{*}$
where $\Phi_s$ has dimensions of the field. The effective expansion
parameter
\begin{equation}
\frac{\overline\lambda}{M} \sim \frac{\lambda^{*}\Phi_s^2}{\sqrt{\lambda^{*}
\Phi^2_s\Phi_m^2}}\label{eq:131}
\end{equation}
There is clearly no particular advantage to not choosing $\Phi_s=\Phi_m$
and one therefore has
\begin{equation}
V_R(\overline\phi) = \frac{\lambda^{*}\Phi_m^2\overline\phi^4}{4!}
+\frac{\lambda^{*\frac{3}{2}}\Phi_m^4}{8\pi\surd 2}\overline\phi^2
-\frac{\lambda^{*\frac{3}{2}}\Phi_m^3}{24\surd 2
\pi}\overline\phi^3\label{eq:132}
\end{equation}
The presence of the two loop logarithm indicates that, in order to
regulate infrared divergences, it will be necessary to scale
$\Phi_m\sim\overline\phi$, giving the leading behaviour
\begin{equation}
V_R \propto
\Bigg(\frac{\lambda^{*}}{4!}+k\lambda^{*\frac{3}{2}}\Bigg)\overline\phi^6\label{eq:133}
\end{equation}
which may be compared with the renormalization group improved potential
given by Lawrie\cite{lawrie2}.

The above technique, quite useful in the scalar theory, is
easily extended to encompass finite temperature states in equilibrium.
Then from the imaginary time formalism one may write\cite{burgess5} in
the massless case
\begin{eqnarray}
V &=& \frac{\overline\lambda+\delta\lambda}{4!}\overline\phi^4
+\frac{1}{2}(\overline m^2+\delta m^2)\overline\phi^2
+\frac{1}{4}\Bigg(\frac{\frac{\overline\lambda}{2}\overline\phi^2
+\delta
m^2_T}{4\pi}\Bigg)^2\Bigg[\ln\Bigg(\frac{\frac{\lambda}{2}\overline\phi^2+\delta  m^2_T}{\mu^2}\Bigg)-\frac{\alpha}{2}\Bigg]\nonumber\\
&+& 2 \beta^{-4}\pi^{\frac{3}{2}}\Gamma(-\frac{3}{2})\sin(\frac{3\pi}{2})
\int_\nu^\infty \frac{(u^2-\nu^2)^{\frac{3}{2}}du}{\exp(2\pi
u)-1}\label{eq:134}
\end{eqnarray}
where $\nu^2=\frac{\beta^2}{4\pi^2}(\frac{\lambda}{2}\overline\phi^2+\delta
m^2_T)$ and $\delta m^2_T$ is the temperature
dependent part of the mass counterterm which is not of order $\hbar$
in the sense of the loop counting parameter and
therefore cannot be neglected. Dependence on $\delta\lambda_T$ cancels in this
scheme. Here it is assumed that the renormalization
is performed at temperature $T=\beta^{-1}$ and not at zero temperature.
Since the renormalization is a resummation of self-energies, it is
natural to implement this at the temperature of interest.
Let
\begin{equation}
I(\nu)=\int_\nu^\infty \frac{(u^2-\nu^2)^{\frac{3}{2}}du}{\exp(2\pi u)-1}.
\label{eq:135}\end{equation}
Appropriate renormalization conditions are now
\begin{eqnarray}
V(0)&=&0\nonumber\\
\frac{\partial^2V}{\partial\overline\phi^2}\Bigg|_{\Phi_R,\beta}&=&\frac{\overline\lambda}{2}\Phi^2_R\nonumber\\
\frac{\partial^4V}{\partial\overline\phi^4}\Bigg|_{\Phi_R,\beta} &=&
\overline\lambda\label{eq:136}
\end{eqnarray}
giving a set of non-linear equations for the counterterms, whose
temperature dependent part must be solved self-consistently. Although there
is no rigid formal distinction between the temperature dependent and
temperature independent counterterms, it is advantageous to keep these
parts separate, since the $\beta$-independent terms contain all the
divergences. One is thus led to define
\begin{eqnarray}
\delta\lambda_T&=&\frac{8}{3}\pi^2\beta^{-4}\frac{\partial^4
I(\nu)}{\partial\overline\phi^4}\Bigg|_{\Phi_R,\beta}\nonumber\\
\delta
m^2_T&=&\frac{8}{3}\pi^2\beta^{-4}\frac{\partial^2I(\nu)}{\partial\overline\phi^2}\Bigg|_{\Phi,\beta} - \frac{\delta\lambda_T}{2}\Phi_R^2.\label{eq:137}
\end{eqnarray}
In a regime where $\overline\phi\sim\Phi_R$, $\delta\lambda_T$ has
no non-trivial effect and only the mass counterterm is important. In a
high temperature regime, it is straightfoward to check that these
equations solve to give the leading order behaviour $\delta m_T^2\sim
\overline\lambda T^2$. This may be pursued in greater detail as
required\cite{elmfors1}.

\section{Scalar Electrodynamics}
Consider now the use of vertex conditions for the
renormalization of the gauge theory. The one loop
effective potential may be computed in an invariant gauge\cite{fradkin1}
provided one uses a Cartesian parameterization of the scalar field.
The results will only be given for $3+1$-dimensions; results in
$2+1$-dimensions have previously been given in\cite{burgess4}. The action
is given by
\begin{equation}
S=\int dV_x\lbrace(D^\mu\Phi)^{\dagger}(D_\mu\Phi)+m^2\Phi^{\dagger}\Phi
+\frac{\lambda}{6}(\Phi^{\dagger}\Phi)^2 -\frac{1}{4}F^{\mu\nu}F_{\mu\nu}
\rbrace\label{eq:138}
\end{equation}
which is to be paramterized in terms of $\Phi=\frac{1}{\surd 2}(\varphi_1
+i\varphi_2)$, or $\varphi_a$ where $a=1,2$.
Consider
the gauge transformation $\delta\varphi^i=R^i_\alpha[\varphi]\delta\xi^\alpha$
in condensed notation, where $\varphi^i = \lbrace
\varphi^a(x),A^\mu(x)\rbrace$.
The components of the Killing vector $R_i[\varphi](x,x')$ may be
found from
\begin{eqnarray}
\delta\varphi_\mu &=& \delta A_\mu = -\partial_\mu\delta\theta(x)\nonumber\\
\delta\varphi_a &=& -e\epsilon_{ab}\varphi_b(x)\delta\theta(x)\label{eq:139}
\end{eqnarray}
where $\theta$ is a gauge transformation. Then
\begin{eqnarray}
R_\mu[\varphi](x,x')&=& -\partial_\mu\delta(x,x')\label{eq:140}\\
R_a[\varphi](x,x')&=& -e\epsilon_{ab}\varphi_b(x')\delta(x,x')\label{eq:141}
\end{eqnarray}
On expanding $\varphi_a = \overline\phi_a +\varphi_a$,
the appropriate gauge fixing condition becomes
$\chi_\alpha=R_{i\alpha}\varphi^i=0$,
or
\begin{equation}
\chi(x) = \partial^\mu A_\mu -
e\epsilon_{ab}\overline\phi_b(x)(\overline\phi_a(x)+\varphi_a(x)) =
0\label{eq:142}
\end{equation}
and the one-loop effective action is given by
\begin{equation}
\Gamma^{(1)}[\overline\phi]= -i\ln\int d\mu[\varphi_a,A_\mu]\Big|{\rm det}
Q\Big|\delta[\partial^\mu A_\mu-e\epsilon_{ab}\varphi_a\overline\phi_b]
e^{\frac{i}{\hbar}S^{(2)}}\label{eq:143}
\end{equation}
where $Q = -\boxx+e^2\overline\phi^2$ and $S^{(2)}$ is the leading part
of the action, quadratic in the quantum fields. Evaluating the
functional integral, one obtains
\begin{eqnarray}
\Gamma^{(1)}&=&\frac{1}{2}{\rm Tr}\ln(-\boxx+m^2+\frac{\lambda}{2})
+\frac{1}{2}{\rm Tr}\ln(-\boxx+e^2\overline\phi^2) + \frac{1}{2}{\rm Tr}\ln A
+\frac{1}{2}{\rm Tr}\ln B\label{eq:144}\\
\lbrace\begin{array}{c}
A\\B
\end{array}\rbrace
&=&
-\boxx+\frac{(m^2+\frac{\lambda}{6}\overline\phi^2+2e^2\overline\phi^2)\pm\sqrt{(m^2+\frac{\lambda}{6}\overline\phi^2+2e^2\overline\phi^2)-4e^4\overline\phi^4}}{2}.\label{eq:145}
\end{eqnarray}
In spontaneous phase transitions one is specifically interested in the
massless limit, whereupon the one loop contribution to the
effective potential becomes
\begin{eqnarray}
64\pi^2\overline\phi^{-4}V^{(1)} &=& e^4\Big[
\ln(\frac{e^2\overline\phi^2}{\mu^2})-\frac{\alpha}{2}\Big]
+\frac{\lambda^2}{4}\Big[
\ln(\frac{\frac{\lambda}{2}\overline\phi^2}{\mu^2})-\frac{\alpha}{2}\Big]\nonumber\\
&+&(\frac{\lambda^2}{72}+\frac{\lambda e^2}{3}+e^4)
\Big[
\ln\left(\frac{\overline\phi^2}{\mu^2}
[\frac{\lambda}{12}+e^2+\frac{1}{2}\sqrt{\frac{\lambda}{6}(\frac{\lambda}{6}+4e^2)}]\right)\nonumber\\
&+&\ln\left(\frac{\overline\phi^2}{\mu^2}[\frac{\lambda}{12}+e^2-\frac{1}{2}\sqrt{\frac{\lambda}{6}(\frac{\lambda}{6}+4e^2)}]\right)
\Big]\nonumber\\
&+&\frac{1}{4}(\frac{\lambda}{6}+2e^2)\sqrt{\frac{\lambda}{6}(\frac{\lambda}{6}+4e^2)}\ln\Bigg[\frac{\frac{\lambda}{6}+2e^2+\sqrt{\frac{\lambda}{6}(\frac{\lambda}{6}+4e^2)}}{\frac{\lambda}{6}+2e^2-\sqrt{\frac{\lambda}{6}(\frac{\lambda}{6}+4e^2)}}\Bigg]\label{eq:146}
\end{eqnarray}
Renormalizing, using the vertex conditions in (\ref{eq:109}-\ref{eq:111}) one
obtains for the
effective potential
\begin{eqnarray}
V_{R}&=&\frac{\overline\lambda}{4!}\overline\phi^4
+\frac{(3e^2+\frac{5}{18}\overline\lambda^2
+\frac{2}{3}\overline\lambda e^2)\overline\phi^4}{64\pi^2}
\Bigg[\ln\Bigg(\frac{\overline\phi^2}{\Phi_\lambda^2}\Bigg)-\frac{25}{6}\Bigg]\nonumber\\
&+&\frac{6\Phi_m^2\overline\phi^2(3e^2+\frac{5}{18}\overline\lambda^2
+\frac{2}{3}\overline\lambda e^2)}{64\pi^2}
\Bigg[\ln\Bigg(\frac{\Phi_\lambda^2}{\Phi_m^2}\Bigg)+3\Bigg]\label{eq:147}
\end{eqnarray}
This is the result obtained by Coleman and Weinberg, up to gauge
dependent terms.
It is noted that, in spite of the gauge sector, the form of the effective
potential is the same as that in the scalar theory, so that the leading
behaviour for small $\overline\phi$ is apparently identical to the scalar
case. The argument for the order of the phase transition also follows
through identically and one concludes, erroneously, that the phase
transition is of second order (continuous). The source of the error lies in
the tacit assumption that the dimensionless quantities $\lambda\sim e^2$.
This arises in neglecting the logarithm of the ratio of the couplings
when going from (\ref{eq:146}) to (\ref{eq:147}) above.
The important point is that the relevant
mass scales are $e^2\overline\phi^2$ and
$\frac{\overline\lambda}{2}\overline\phi^2$ and not merely $\overline\phi^2$.
The vertex condition (\ref{eq:109}), insensitively applied, gives preference to
the
scalar mass at the expense of the effective working gauge mass.
The same problem occurs in the large-$N$ expansion in which the gauge
contribution is suppressed by $1/N$.

It should be remarked that the correct result can be obtained from the
renormalization group provided one includes counterterms for the
electric charge and for a background gauge field. Although these are
formally zero in the zero momentum limit, the additional parametric
dependence on the gauge sector prevents the coarse mass renormalization
from washing out the effect of the gauge fields. One wonders, on the
other hand, why it is necessary to work at non-zero momentum to
obtain a zero momentum result. The answer is simply that one needs to
renormalize the gauge sector on an equal basis with the scalar sector,
but that there is no natural way to do this at zero momentum using
vertex conditions as a renormalization scheme, since the beta function
for the electric charge vanishes in this limit.
It is sufficient to compute the effective action to quadratic order
in the background fields $\overline\phi$, $\overline A_\mu$. Defining
for the purposes of minimal subtraction
$\overline\phi_B=Z_\phi^{\frac{1}{2}}\overline\phi$ and
$\overline A^\mu_B=Z_\phi^{\frac{1}{2}}\overline A^\mu$,
and additive counterterms for the remainder one can after
lengthy calculation show that
\begin{eqnarray}
\Gamma^{(1)}_{quad} &=& \int dV_x
\Big\lbrace
e^2\overline A^\mu \overline A_\mu \Delta(x,x) + \frac{1}{2}e^2\overline\phi^2
G^\mu_{~\mu}(x,x) + \frac{\lambda}{3}\overline\phi^2\Delta(x,x)
+\frac{1}{2\alpha}
\Big\rbrace\nonumber\\
&-&\frac{e^2}{2}\int dV_x\,dV_{x'}
\Big\lbrace
2 \overline A^\mu \overline A_\mu\,\Delta(x,x')\stackrel{x}{\partial_\mu}
\stackrel{x'}{\partial_\nu}\Delta(x,x')
-2\overline A^\mu \overline A^\nu\; \stackrel{x}{\partial_\mu}\Delta(x,x')
\stackrel{x'}{\partial_\nu}\Delta(x,x') \nonumber\\
&+&
\overline\phi^2
\,G^{\mu\nu}(x,x')\stackrel{x}{\partial_\mu}\stackrel{x'}{\partial_\nu}\Delta(x,x')
+(\stackrel{x}{\partial_\mu}\overline\phi)(\stackrel{x'}{\partial_\nu}\overline\phi) G^{\mu\nu}(x,x')\Delta(x,x')\nonumber\\
&+&\frac{1}{\alpha}\overline\phi^2
\Delta(x,x')\stackrel{x}{\partial_\mu}\stackrel{x'}{\partial_\nu}G^{\mu\nu}(x,x)
-2\overline\phi(\stackrel{x'}{\partial_\nu}\overline\phi)
G^{\mu\nu}(x,x')\stackrel{x}{\partial_\mu}\Delta(x,x')\nonumber\\
&+&\frac{2}{\alpha}\overline\phi^2 \stackrel{x'}{\partial_\nu} G^{\mu\nu}(x,x')
\stackrel{x}{\partial_\mu}\Delta(x,x')
-\frac{2}{\alpha}(\stackrel{x}{\partial_\mu}\overline\phi)\overline\phi(x')
\stackrel{x'}{\partial_\nu}G^{\mu\nu}(x,x')\Delta(x,x')
\Big\rbrace
\end{eqnarray}
where
\begin{eqnarray}
(-\boxx+m^2)\Delta(x,x')=\delta(x,x')\nonumber\\
(-\boxx\delta^\mu_{~\nu}+ (1-\frac{1}{\alpha})\partial^\mu\partial_\nu)
G^\nu_{~\lambda}=\delta^\mu_{~\lambda}\delta(x,x')
\end{eqnarray}
and the pole-parts of the above products of Green
functions can be extracted using dimensional regularization.
See for instance the method used in \cite{parker1}.
It is noted briefly that terms varying like negative powers of $\alpha$
cancel as they must for gauge fixing independence in the Landau-DeWitt
gauge ($\alpha\rightarrow 0$) and terms quadratic in the background
photon field cancel, confirming renormalizability.
The computations are rather lengthy and will not be given here. On comparison
to the quadratic counterterm action,
\begin{eqnarray}
S_{CT}^{(1)}=\int dV_x \Big\lbrace
\frac{1}{2}\delta Z_\phi\,\overline\phi(-\boxx)\overline\phi
+\frac{1}{2}(\delta m^2 + m^2\delta Z_\phi)\overline\phi^2
+\frac{1}{2}\delta Z_A\overline A^\mu (-\boxx\delta_{\mu\nu}
-\partial_\mu\partial_\nu ) \overline A^\nu
\Big\rbrace
\end{eqnarray}
and using the relation $(e+\delta e)\delta Z_A^{\frac{1}{2}}=e$, one
obtains the one-loop counterterms for the gauge theory.
\begin{eqnarray}
\delta Z_\phi &=& -6e^2\epsilon^{-1}(4\pi)^{-2}\nonumber\\
\delta Z_A &=&\frac{2}{3}e^2\epsilon^{-1}(4\pi)^{-2}\nonumber\\
\delta e &=& -\frac{1}{3}e^3\epsilon^{-1}(4\pi)^{-2}\nonumber\\
\delta m^2 &=& = (2 m^2e^2-\frac{4}{3}\lambda m^2)
\epsilon^{-1}(4\pi)^{-2}\nonumber\\
\delta\lambda &=& -12(\frac{5}{18}\lambda^2+3e^4-\frac{4}{3}\lambda
e^2)\epsilon^{-1}(4\pi)^{-2}
\end{eqnarray}
where $\epsilon=n-4$. Thus there are non-zero counterterms for the gauge sector
which will lead to a renormalization group flow.
The move to non-zero
momentum can be construed as artifical; a better approach would be
desirable. One can speculate as to whether a different scheme for
finding beta functions at zero momentum would result in the appearence
of fixed points associated with a first order transition.

\bibliographystyle{unsrt}

\end{document}